\makeatletter
\@ifundefined{@parse@version@dash}{%
\def\@parse@version#1{\@parse@version@0#1}
\def\@parse@version@#1/#2/#3#4#5\@nil{%
\@parse@version@dash#1-#2-#3#4\@nil}
\def\@parse@version@dash#1-#2-#3#4#5\@nil{%
  \if\relax#2\relax\else#1\fi#2#3#4 }
}{}
\makeatother

\documentclass[%
 reprint,
superscriptaddress,
nofootinbib,
 amsmath,amssymb,
 aps,
 prd,
floatfix
]{revtex4-2}

\usepackage{graphicx}
\usepackage{dcolumn}
\usepackage{bm}

\usepackage{combelow}
\usepackage{color}
\usepackage{xcolor}  
\usepackage{braket}

\usepackage[normalem]{ulem}

\def\hati{{\hat{\imath}}}
\def\halpha{{\hat{\alpha}}}
\def\hbeta{{\hat{\beta}}}
\def\hgamma{{\hat{\gamma}}}
\def\hsigma{{\hat{\sigma}}}
\def\htau{{\hat{\tau}}}
\def\hatr{{\hat{r}}}
\def\htheta{{\hat{\theta}}}
\def\heta{{\hat{\eta}}}

\def\d{{\mathrm{d}}}

\def\wk{{\overline{k}}}

\begin{document}

\title{Bjorken flow attractors with transverse dynamics}%

\author{Victor E. Ambru\cb{s}}
\thanks{Corresponding author}
\email{victor.ambrus@e-uvt.ro}
\affiliation{Institut f\"ur Theoretische Physik, Johann Wolfgang Goethe-Universit\"at, Max-von-Laue-Strasse 1, D-60438 Frankfurt am Main, Germany}%
\affiliation{Department of Physics, West University of Timi\cb{s}oara, \\
Bd.~Vasile P\^arvan 4, Timi\cb{s}oara 300223, Romania}

\author{Sergiu Busuioc}
\email{sergiu.busuioc@ed.ac.uk}
\affiliation{School of Engineering, University of Edinburgh, Edinburgh, EH9 3FB, United Kingdom}

\author{Jan A. Fotakis}%
\email{fotakis@itp.uni-frankfurt.de}
\affiliation{Institut f\"ur Theoretische Physik, Johann Wolfgang Goethe-Universit\"at, Max-von-Laue-Strasse 1, D-60438 Frankfurt am Main, Germany}%

\author{Kai Gallmeister}
\email{gallmeister@itp.uni-frankfurt.de}
\altaffiliation[\it Present address:]{Institut f\"ur Theoretische Physik, Justus-Liebig-Universit\"at, Heinrich-Buff-Ring 16, 35392 Gie{\ss}en, Germany}
\affiliation{Institut f\"ur Theoretische Physik, Johann Wolfgang Goethe-Universit\"at, Max-von-Laue-Strasse 1, D-60438 Frankfurt am Main, Germany}%

\author{Carsten Greiner}
\email{carsten.greiner@itp.uni-frankfurt.de}
\affiliation{Institut f\"ur Theoretische Physik, Johann Wolfgang Goethe-Universit\"at, Max-von-Laue-Strasse 1, D-60438 Frankfurt am Main, Germany}%

\date{\today}

\begin{abstract}
In the context of the longitudinally boost-invariant
Bjorken flow with transverse expansion, we use three 
different numerical methods to analyze the emergence 
of attractor solutions in an ideal gas of massless
particles exhibiting constant shear viscosity to entropy
density ratio $\eta / s$. The fluid energy density is
initialized using a Gaussian profile in the transverse
plane, while the ratio $\chi = \mathcal{P}_L /
\mathcal{P}_T$ between the longitudinal and transverse
pressures is set at initial time $\tau_0$ to a constant
value $\chi_0$ throughout the system employing the
Romatschke-Strickland distribution. We introduce the
hydrodynamization time $\delta \tau_H = (\tau_H -
\tau_0)/ \tau_0$ based on the time $\tau_H$ when the
standard deviation $\sigma(\chi)$ of a family of
solutions with different $\chi_0$ reaches a 
minimum value at the point of maximum convergence 
of the solutions. In the $0+1{\rm D}$ setup, 
$\delta \tau_H$ exhibits scale invariance, being a
function only of $(\eta / s) / (\tau_0 T_0)$. 
With transverse expansion, we find a similar 
$\delta \tau_H$ computed with respect to the 
local initial temperature, $T_0(r)$. We highlight 
the transition between the regimes where the
longitudinal and transverse expansions dominate. 
We find that the hydrodynamization time required 
for the attractor solution to be reached increases 
with the distance from the origin, as expected based 
on the properties of the $0+1{\rm D}$ system defined 
by the local initial conditions. We argue that
hydrodynamization is predominantly the effect of 
the longitudinal expansion, being significantly 
influenced by the transverse dynamics only for 
small systems or for large values of $\eta / s$.
\end{abstract}

\maketitle


\section{Introduction} \label{sec:intro}

The Bjorken model for a longitudinally boost-invariant expanding system \cite{Bjorken:1982qr} has proven successful for the description of the fluid phase of the quark-gluon plasma created after the collision of highly-energetic ultrarelativistic heavy ions \cite{Romatschke:2007mq,Weller:2017tsr}. 

In the context of the transversally-homogeneous Bjorken expansion (called the $0+1{\rm D}$ Bjorken flow), it was shown that the information regarding the nonequilibrium state of the system (i.e., the ratio $\chi = \mathcal{P}_L / \mathcal{P}_T$ between the longitudinal and transverse pressures) disappears after a finite timescale (called the {\it hydrodynamization timescale} \cite{Heller:2015dha}).
In the early onset of the rapid longitudinal expansion, the momentum distribution of the partons is strongly transversal \cite{Mueller:1999pi}, before the counter balancing of the dissipative impact of collisions takes over to distribute the momenta in the longitudinal direction as the Bjorken expansion time increases \cite{Mueller:1999pi,El:2009vj}.
In this still early regime, attractor solutions can develop, which were shown to exist for a wide class of fluids (e.g., hard spheres \cite{Denicol:2019lio} and constant shear viscosity to entropy density $\eta / s$ ratio \cite{Heller:2015dha}), by using a variety of off-equilibrium models, such as hydrodynamics \cite{Heller:2015dha,Denicol:2019lio},  conformal \cite{Heller:2016rtz} and nonconformal \cite{Romatschke:2017acs} kinetic theory, the Fokker-Planck model for gluons \cite{Behtash:2020vqk}, $\mathcal{N}=4$ SYM model for strongly-coupled plasmas \cite{Kurkela:2019set,Romatschke:2018qwe} or the effective kinetic theory (EKT) for weakly coupled QCD \cite{Almaalol:2020rnu}. In the context of the Gubser model, which accounts for transverse expansion via the Gubser symmetry group \cite{Gubser:2010ze}, the existence of attractor solutions has been considered in Refs.~\cite{Behtash:2019qtk,Dash:2020zqx}.

As pointed out in Ref.~\cite{Heller:2020anv}, in more realistic systems, the attractor behavior may be observed for quantities which differ from the pressure anisotropy denoted in the present work by $\chi$. In such cases, it is instructive to search for the attractor behavior at the level of the phase space. In this work, we focus on systems exhibiting longitudinal boost invariance which are nearly conformal, where the pressure anisotropy $\chi$ provides a good measure of hydrodynamization.

As discussed in Ref.~\cite{Romatschke:2017acs} in the context of the resummed Baier-Romatschke-Son-Starinets-Stephanov
\cite{Baier:2007ix} (rBRSSS) theory, the attractor solutions can be identified also in systems with transverse expansion. In Ref.~\cite{Kurkela:2020wwb}, the properties of elliptic flow in Bjorken-like systems with transverse expansion were investigated from the perspective of the early-time attractor of the underlying $0+1{\rm D}$ Bjorken flow. As pointed out in Ref.~\cite{Kurkela:2019set}, hydrodynamization in systems with transverse dynamics may be expected to occur as in the equivalent $0+1{\rm D}$ setup when the transverse gradients are weaker than the corresponding longitudinal ones. Our present work reasserts this expectation by considering finite-size systems corresponding to $p$-$p$, $p$-$A$ (small) or $A$-$A$ (large) collisions. 

In this paper, we take the approach of characterizing 
the onset of hydrodynamization on the basis of the loss
of memory with respect to the initial pressure anisotropy 
$\chi_0$. For this purpose, we
consider a family of systems initialized with various
values of $\chi_0$ and compute, at each temporal
instance $\tau$ (and each radial distance $r$ for 
the systems with transverse expansion), the standard
deviation $\sigma(\chi)$ of the pressure anisotropy,
taken with respect to the $\chi_0$ ensemble. As the
hydrodynamic attractor is approached, the curves
corresponding to these systems converge toward each
other, causing $\sigma(\chi)$ to decrease. We consider
that hydrodynamization is achieved at the time 
$\tau_H$ when $\sigma(\chi)$ reaches its minimum 
value $\sigma_{\rm min}$ corresponding to the point 
of maximal convergence. This value is not strictly 
zero for two reasons, which we investigate in this
paper. The first reason concerns the time frame at 
which the curves corresponding to various values of 
$\chi_0$ intersect each other, which has a small 
but finite temporal extent. The second reason why 
$\sigma(\chi)$ stays finite is that, after 
$\sigma(\chi)$ reaches its minimum, the family 
of solutions overshoots past the convergence point.
This overshoot leads for a short time to an increase 
of $\sigma(\chi)$, after which $\sigma(\chi)$
resumes its decreasing trend, confirming the 
validity of the attractor solution.

For practical applications, one can consider that 
the system loses the memory regarding its initial 
state when $\sigma(\chi)$ drops below a certain
threshold value $\sigma_{\rm th}$ (or when it 
reaches the minimum value $\sigma_{\rm min}$, 
if this value is larger than $\sigma_{\rm th}$). 
The threshold can be regarded as a free-streaming
regulator, when $\sigma_{\rm min} = 0$ is reached 
only asymptotically as $\tau \rightarrow \infty$.
We quantify the efficacy of hydrodynamization on 
the basis of the hydrodynamization timescale 
$\delta \tau_H^{\sigma_{\rm th}} = 
(\tau_H^{\sigma_{\rm th}} - \tau_0) / \tau_0$, 
where $\tau_H^{\sigma_{\rm th}}$ and $\tau_0$ 
are the values of the time coordinate when the
hydrodynamization criterion is reached and at
initialization, respectively.
In Sec.~\ref{sec:dtauH}, we reveal that in the 
$0+1{\rm D}$ boost-invariant setup, 
$\delta \tau_H$ is a function only of the 
combination $(\eta / s) / (\tau_0 T_0)$. 

The paper is structured as follows. In
Sec.~\ref{sec:bjork}, we review the $0+1{\rm D}$ 
Bjorken flow setup. The hydrodynamization process 
is investigated using three different methods, 
namely: second order hydrodynamics, Boltzmann 
approach to multi-parton scattering and the 
relaxation time approximation of the relativistic
Boltzmann equation. In Sec.~\ref{sec:dtauH}, we
introduce the hydrodynamization timescale 
$\delta \tau_H$ and discuss its scaling properties 
in the $0+1{\rm D}$ setup. In Sec.~\ref{sec:transv}, 
we investigate the hydrodynamization in systems 
with transverse expansion and discuss the 
consequences of transverse expansion on the
hydrodynamization timescale $\delta \tau_H$. Our
conclusions are summarized in Sec.~\ref{sec:conc}. 
This paper is supplemented by two Appendices. 
In Appendix~\ref{app:HS}, we address the 
$0+1{\rm D}$ Bjorken flow for hard-sphere particles
within the three frameworks mentioned above.
Appendix~\ref{app:RTA} presents a brief description 
of the RTA numerical method.

\section{$0+1{\rm D}$ Bjorken flow}\label{sec:bjork}

We begin our analysis by revisiting the $0+1{\rm D}$
Bjorken flow with full transverse plane homogeneity. 
Here and henceforth, we restrict our analysis to the
case of an ultrarelativistic gas of massless particles,
for which the energy density $e$ and isotropic pressure
$p$ are related via $e = 3p$. In order to take advantage
of the longitudinal boost-invariance, it is convenient
to work with the Bjorken time $\tau = \sqrt{t^2 - z^2}$
and space-time rapidity 
$\eta_s = \frac{1}{2} \ln \frac{t + z}{t - z}$, 
giving rise to the line element
\begin{equation}
 \d s^2 = \d\tau^2 - \d x^2 -\d y^2 - \tau^2 \d\eta_s^2.
 \label{eq:ds2}
\end{equation}
The conservation of the energy-momentum tensor, 
$\nabla_\nu T^{\mu\nu} = 0$, entails
\begin{equation}
 3 \tau \partial_\tau p + 4p + \pi = 0,
 \label{eq:1D_cons}
\end{equation} 
where $\pi$ is a measure of the pressure anisotropy
which can be related to the longitudinal 
($\mathcal{P}_L$) and transverse ($\mathcal{P}_T$)
pressures via
\begin{equation}
 \mathcal{P}_L = p + \pi, \qquad 
 \mathcal{P}_T = p - \frac{\pi}{2}. 
 \label{eq:PLPT}
\end{equation}
The time evolution of $\pi$ must be supplied by an
equation which is highly dependent on the model 
employed for the description of the system. In this
work, we consider three methods to compute the solution
of the above equation, namely the viscous  SHArp and
Smooth Transport Algorithm (vSHASTA) 
\cite{Molnar:2009tx,Niemi:2012ry,Fotakis:2019nbq} 
for relativistic hydrodynamics ({\it hydro}), 
the lattice Boltzmann method 
\cite{Romatschke:2011hm,Ambrus:2018kug,Gabbana:2019ydb}
for the relativistic Boltzmann equation in the 
Anderson-Witting relaxation time approximation for the 
collision term \cite{Anderson:1974,Anderson:1974b}
(RTA), and the Boltzmann Approach to Multi-Parton
Scattering \cite{Xu:2004mz,Xu:2007jv} (BAMPS). 

The RTA numerical solver is based on the vielbein
formalism, extending the implementation in
Ref.~\cite{Ambrus:2018kug} to take into account the
azimuthally symmetric flow in the transverse plane. 
The details regarding this extension are presented 
in Appendix~\ref{app:RTA}. The BAMPS results shown 
in this work are generated with an optimized code
version, which still works in 3D Cartesian space
coordinates, but makes use of the longitudinal boost
invariance. Since thus only particles in the 
transversal plane at midrapidity have to be 
considered, numerical statistics better than 
$10^5$ compared to the calculations in
\cite{Gallmeister:2018mcn} is possible.

Since BAMPS is a particle-based solver, it 
automatically conserves the particle four-flow 
$N^\mu$ when only elastic binary collisions are 
taken into account \cite{El:2009vj,Gallmeister:2018mcn}.
Therefore, Eq.~\eqref{eq:1D_cons} is supplemented by 
the condition $\partial_\mu N^\mu = 0$, which reduces 
in the case of the $0+1{\rm D}$ Bjorken flow to
\cite{Denicol:2019lio}
\begin{equation}
 \partial_\tau(n \tau) = 0 \quad \Rightarrow \quad
 n(\tau) = \frac{n_0 \tau_0}{\tau},
\end{equation}
where $n$ is the particle number density and $\tau_0$ 
is the initial time. In the theory of second-order
hydrodynamics derived based on the 14-moment
approximation in the context of the Anderson-Witting
model, $\pi$ satisfies the following evolution 
equation \cite{Jaiswal:2013npa}:
\begin{equation}
 \frac{\partial \pi}{\partial \tau} = 
 -\frac{\pi}{\tau_R} - \beta_\pi \frac{4}{3\tau} 
 - \lambda \frac{\pi}{\tau},
 \label{eq:1D_Pi}
\end{equation}
where for a system consisting of a massless Boltzmann
gas, we have $\beta_\pi = \eta / \tau_R$ and 
$\lambda = 38/21$ \cite{Jaiswal:2013npa}. The 
relaxation time $\tau_R$ is related to the shear
viscosity via \cite{Cercignani02}
\begin{equation}
 \eta = \frac{4}{5} \tau_R p.
 \label{eq:RTA_eta_tau}
\end{equation}
The initial pressure anisotropy ratio 
$\chi_0 \equiv \mathcal{P}_L(\tau_0) / 
\mathcal{P}_T(\tau_0)$ is introduced through the 
initial choice of $\pi$ via Eqs. \eqref{eq:PLPT},
as follows:
\begin{equation}
 \pi_0 = -p_0\frac{1 - \chi_0}{1 + \chi_0 / 2},
\end{equation}
where $p_0 = e_0 / 3$ is the pressure at initial 
Bjorken time $\tau_0$. 

In the RTA and BAMPS approaches, the initial pressure
anisotropy is modeled by setting $f$ to be equal to 
the Romatschke-Strickland distribution for the ideal 
gas \cite{Romatschke:2003ms,Florkowski:2013lya},
\begin{equation}
 f_{\rm RS} = \frac{g e^{\alpha_0}}{(2\pi)^3} 
 \exp \left[-\frac{1}{\Lambda_0} \sqrt{(k \cdot u)^2 
 + \xi_0 (k \cdot \hat{z})^2}\right],
 \label{eq:RS_gen}
\end{equation}
where $k^\mu$ and $u^\mu$ are the particle momentum and
macroscopic velocity four-vectors, while $\hat{z}^\mu$
is the unit-vector along the rapidity coordinate. 
With respect to the Bjorken coordinates, $u^\mu$ and 
$\hat{z}^\mu$ have only one nonvanishing component,
i.e. $u^\tau = 1$ and $\hat{z}^{\eta_s} = \tau^{-1}$.
Expressing the momentum vector $k^\mu$ in terms of 
$k$, $\xi$ and $\varphi$ defined via
\begin{equation}
 k^\tau = k, \quad 
 \begin{pmatrix}
  k^x \\ k^y
 \end{pmatrix} = k\sqrt{1 - \xi^2}
 \begin{pmatrix}
  \cos\varphi \\ \sin\varphi
 \end{pmatrix}, \quad 
 k^{\eta_s} = \frac{k \xi}{\tau},
\end{equation}
Eq.~\eqref{eq:RS_gen} reduces to
\begin{equation}
 f_{\rm RS} = \frac{g e^{\alpha_0}}{(2\pi)^3} 
 \exp \left(-\frac{k}{\Lambda_0} 
 \sqrt{1 + \xi_0 \xi^2}\right).
 \label{eq:RS}
\end{equation}
The degeneracy is set to $g = 16$ to account for the
gluonic degrees of freedom. The anisotropy parameter 
$\xi_0$ takes the value $0$ for an isotropic 
(Maxwell-J\"uttner) distribution and $\infty$ for 
an infinitely skewed distribution. The parameters 
$\alpha_0$ and $\Lambda_0$ allow the initial 
particle number density and pressure to be 
specified independently via
\begin{align}
 e^{\alpha_0} =& \frac{\pi^2 n_0}{g \Lambda_0^3} 
 \sqrt{1 + \xi_0}, \nonumber\\
 \Lambda_0 =& \frac{2p_0 /n_0}{\sqrt{1 + \xi_0}}
 \left(\frac{\arctan\sqrt{\xi_0}}{\sqrt{\xi_0}} 
 + \frac{1}{1 + \xi_0}\right)^{-1}.
 \label{eq:RS_aux}
\end{align}
In this work, we consider that at initial time, the
chemical potential vanishes, such that 
$n_0 = g T_0^3 / \pi^2$. The initial longitudinal 
and transverse pressures are \cite{Ambrus:2019wfc}
\begin{align}
 \mathcal{P}_{L;0} =& \frac{3 g \Lambda_0^4 e^{\alpha_0}}{2\pi^2 \xi_0} 
 \left(\frac{\arctan \sqrt{\xi_0}}{\sqrt{\xi_0}} - \frac{1}{1 + \xi_0}\right),\nonumber\\
 \mathcal{P}_{T;0} =& \frac{3 g \Lambda_0^4 e^{\alpha_0}}{4\pi^2 \xi_0} 
 \left[1 + (\xi_0 - 1) \frac{\arctan \sqrt{\xi_0}}{\sqrt{\xi_0}}\right],
\end{align}
such that their ratio 
$\chi_0 = \mathcal{P}_{L;0} / \mathcal{P}_{T;0}$ 
depends solely on the parameter $\xi_0$:
\begin{equation}
 \chi_0 = \frac{2}{1 + \xi_0} \frac{(1 + \xi_0) \frac{\arctan\sqrt{\xi_0}}{\sqrt{\xi_0}} - 1}
 {1 + (\xi_0 - 1) \frac{\arctan\sqrt{\xi_0}}{\sqrt{\xi_0}}}.
 \label{eq:RS_aux2}
\end{equation}
Negative values of $\xi_0$, corresponding to 
$\chi_0 > 1$, are not considered in this paper. 
The details regarding the RTA solver used in the
$0+1{\rm D}$ case were given in 
Refs.~\cite{Ambrus:2018kug,Ambrus:2019wfc} and 
are summarized in Appendix~\ref{app:RTA}.

\begin{figure}
\begin{tabular}{c}
\includegraphics[width=0.95\columnwidth]{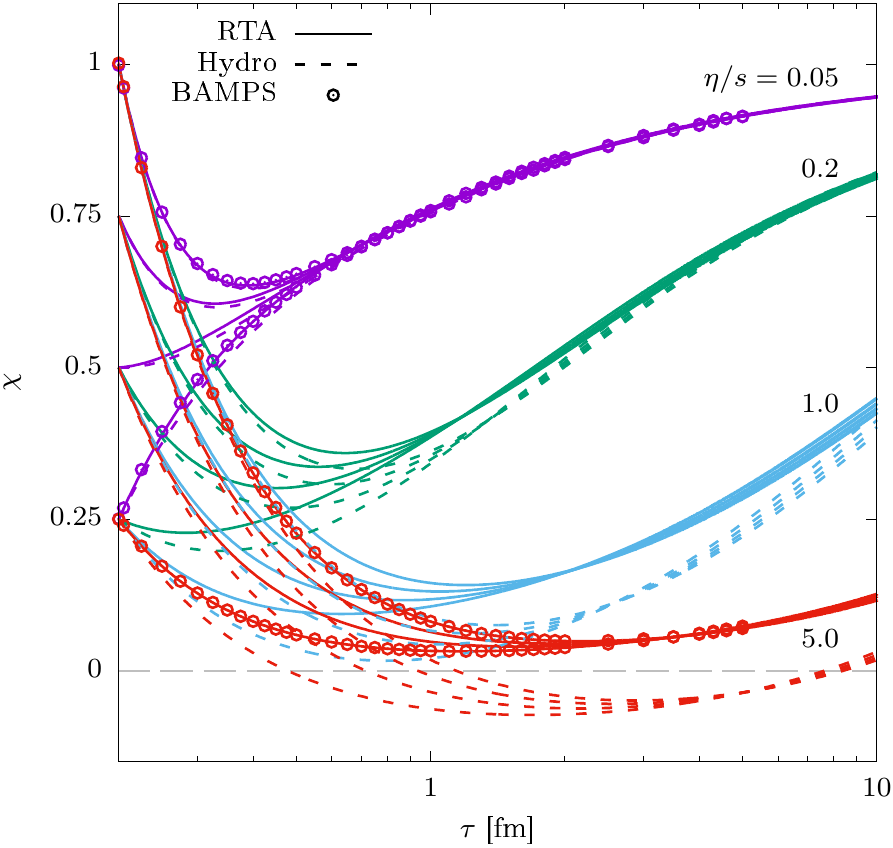}
\end{tabular}
\caption{
Evolution of the pressure anisotropy 
$\chi = \mathcal{P}_L / \mathcal{P}_T$ with respect 
to the Bjorken time $\tau$. The RTA and hydro 
results are shown with solid and dashed lines,
respectively, while the BAMPS results are shown 
using empty circles.
\label{fig:1D_profiles}
}
\end{figure}

Figure~\ref{fig:1D_profiles} shows a comparison between
the results obtained using the three methods enumerated
above for $\eta / s = 0.05$, $0.2$, $1$ and $5$. 
The initial time (here and henceforth, unless otherwise
specified) is set to $\tau_0 = 0.2\ {\rm fm}$ and the 
initial temperature is set to $T_0 = 0.5\ {\rm GeV}$.
The anisotropy parameter is taken such that 
$\chi_0 \in \{0.25, 0.5, 0.75, 1\}$. At small 
$\eta / s$, all methods are in very good agreement 
with each other. At large $\eta / s$, the RTA and 
BAMPS maintain agreement, while the hydro results
present significant deviations (see in this respect
Ref.~\cite{El:2009vj}). In particular, the hydro 
results achieve negative values for $\chi$ at 
$\eta /s = 5$, signaling the breakdown of the 
hydrodynamic equations in this regime. Possible
resolutions to this problem include third order
extensions of hydrodynamics 
\cite{El:2009vj,Jaiswal:2013vta} and the 
anisotropic hydrodynamics framework 
\cite{Florkowski:2010cf,Martinez:2010sc,Molnar:2016gwq},
however we do not pursue this further in what follows. 

In addition, a comparison between our three numerical
methods in the case of a hard-sphere gas (interacting
via a constant cross-section) is presented in
Appendix~\ref{app:HS}.

\section{Hydrodynamization timescale $\delta \tau_H$}
\label{sec:dtauH}

By looking at Fig.~\ref{fig:1D_profiles}, it is obvious
that the curves corresponding to different initial 
anisotropies $\chi_0$ merge after some time $\tau_H$,
which increases with $\eta / s$. This can happen either 
due to the approach to the attractor solution or due to
a ``memory-loss process'' which effectively causes all
curves to collapse on top of each other (see, e.g., 
the free streaming limit discussed below). Also, 
the merger time can be seen to be larger for the 
hydro curves than for the kinetic theory curves 
(RTA and BAMPS). Without making any prior assumption 
about the mathematical nature (or even existence) of 
a universal attractor solution for this type of flow, 
we characterize the efficacy of hydrodynamization 
based on the hydrodynamization timescale 
$\delta \tau_H = (\tau_H - \tau_0) / \tau_0$ on which
the solution becomes independent of the initial pressure
anisotropy. The quantity $\delta \tau_H$ is introduced
formally in Sec.~\ref{sec:dtauH:def} and its behavior 
at small $(\eta / s)/(\tau_0 T_0)$ is discussed in 
Sec.~\ref{sec:dtauH:trans} on the basis of a 
transseries representation of $\chi$. 
Its properties in the extreme case of a free streaming fluid are
considered in Secs~\ref{sec:dtauH:FS_hydro} and
\ref{sec:dtauH:FS_RTA} for hydro and RTA, respectively.
The scaling properties of $\delta \tau_H$ at finite 
relaxation time are discussed in 
Sec.~\ref{sec:dtauH:res}.

\subsection{Definition}\label{sec:dtauH:def}

\begin{figure}
\begin{tabular}{c}
\includegraphics[width=0.95\columnwidth]{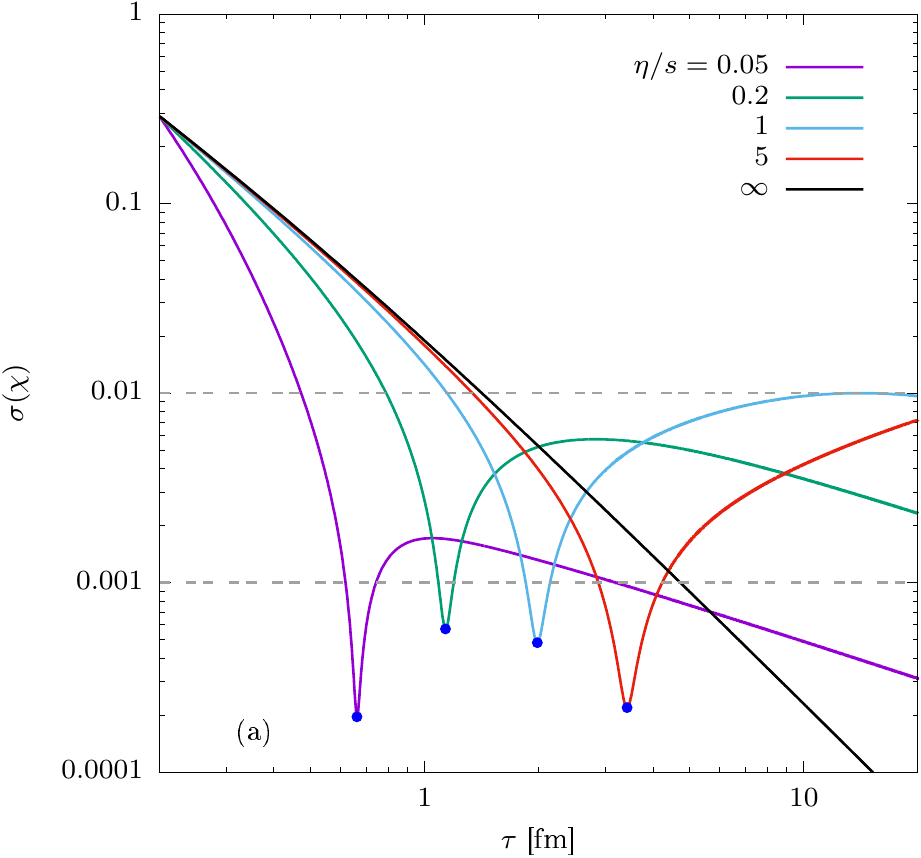} \\
\includegraphics[width=0.95\columnwidth]{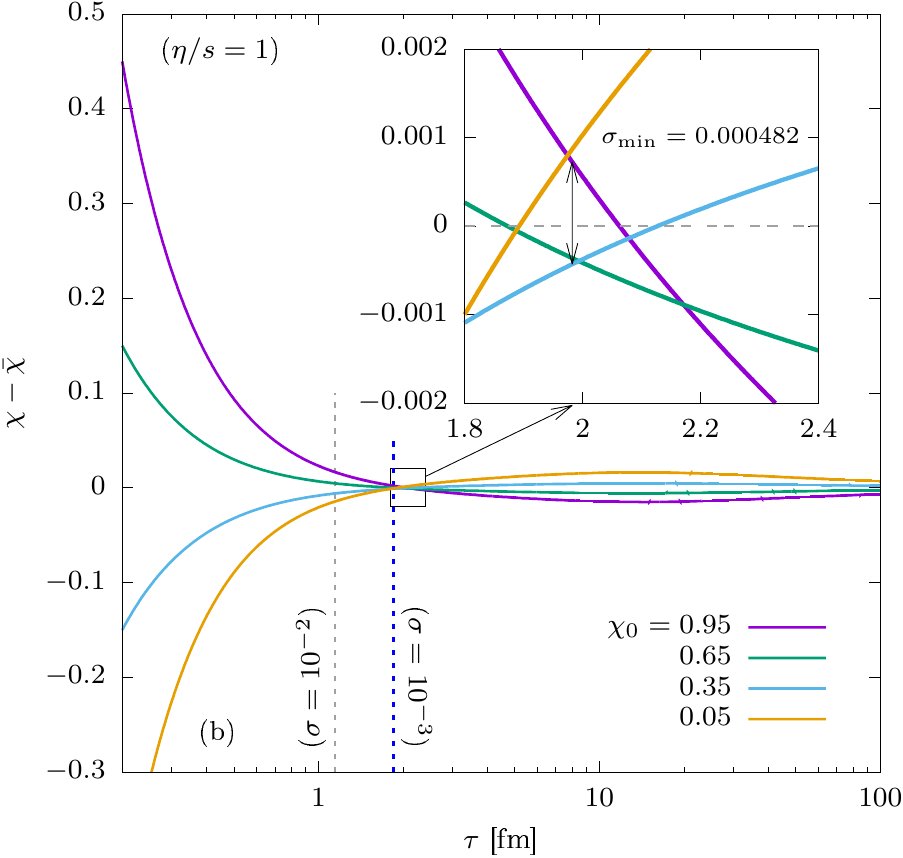}
\end{tabular}
\caption{
RTA results for (a) the dependence of $\sigma(\chi)$ on $\tau$ for various values of $\eta / s$; and (b) the dependence of $\chi - {\bar \chi}$ on $\tau$ for various values of $\chi_0$ at $\eta / s = 1$.
\label{fig:1D_sigma}
}
\end{figure}

Quantitatively, the memory-loss effect can be assessed
by looking at the standard deviation $\sigma$ of $\chi$
with respect to the initial pressure ratio $\chi_0$,
\begin{equation}
 \sigma(\chi) = \left[\int_0^1 \d\chi_0 
 (\chi - \overline{\chi})^2\right]^{1/2}, \qquad 
 \overline{\chi} = \int_0^1 \d\chi_0\, \chi.
 \label{eq:sigma_def}
\end{equation}
The details regarding the computation of $\sigma(\chi)$
and $\overline{\chi}$ from the simulation data are 
given at the end of Sec.~\ref{sec:dtauH:res}.
The time dependence of $\sigma(\chi)$ computed 
within the RTA framework is shown in 
Fig.~\ref{fig:1D_sigma}(a) for the four cases 
considered in Fig.~\ref{fig:1D_profiles}, as well as 
for the free-streaming (FS) regime 
($\eta / s \rightarrow \infty$), which will be 
discussed in Sec.~\ref{sec:dtauH:FS_RTA}. 
In the FS regime, $\sigma(\chi)$ decreases 
monotonously with $\tau$. For finite $\eta / s$, 
$\sigma(\chi)$ exhibits a rebound after it reaches 
a minimum (but very small) value (indicated by the 
blue dots) $\sigma_{\rm min} \equiv 
\sigma_{\rm min}(\eta / s)$, which depends on the 
value of $\eta / s$. A nonmonotonic behavior of 
this minimum value can be seen, being lower for 
small ($0.05$) and large ($5$) values of $\eta / s$, 
and larger for the intermediate values ($0.2$ and $1$).
After this rebound, a tail of milder descending slope 
is observed, leading to smaller values of 
$\sigma(\chi)$ as $\tau \rightarrow \infty$.

The nature of the minimum marked by the blue dots can 
be understood already from Fig.~\ref{fig:1D_profiles}.
It can be seen that, after the $\chi$ curves for a 
given value of $\eta / s$ corresponding to various
values of $\chi_0$ intersect, they have a tendency 
to overshoot. This tendency is more pronounced for 
$\eta / s = 0.2$ and $1$, which is consistent with 
the results for $\sigma(\chi)$ seen above. Further
details can be seen by looking at the time evolution 
of $\chi - {\bar \chi}$, shown for $\eta /s = 1$ in
Fig.~\ref{fig:1D_sigma}(b). After intersection, the
lines corresponding to different values of $\chi_0$ 
tend to follow a tubelike trajectory of finite width,
which eventually decreases as $\tau \rightarrow \infty$.
The inset shows that the curves corresponding to 
various initial values of $\chi_0$ intersect the 
curve corresponding to $\chi = {\bar \chi}$ at 
different times, causing $\sigma$ to remain finite
throughout the entire hydrodynamization process. 
The minimum value of $\sigma(\chi)$ for $\eta / s = 1$
is $\sigma_{\rm min} = 4.82 \times 10^{-4}$, which is
indeed very small, but finite. The times when 
$\sigma(\chi)$ drops below $10^{-2}$ and $10^{-3}$ 
are shown by the vertical dotted lines in the main 
plot.

The discussion above prompts us to characterize 
the progression of the hydrodynamization process 
from the perspective of 
$\sigma(\chi) \equiv \sigma(\chi; \tau)$. 
We consider that the system reaches hydrodynamization 
at $\tau_H^0$ when $\sigma(\chi; \tau_H^0)$ reaches the
minimum value $\sigma_{\rm min}$ ($\tau_H^0$ is about
$0.66$, $1.14$, $1.98$ and $3.42\ {\rm fm}$ for 
$\eta / s = 0.05$, $0.2$, $1$ and $5$, respectively).
The hydrodynamization timescale in this case is 
denoted $\delta \tau_H^0$. From 
Fig.~\ref{fig:1D_sigma}(a), it can be expected that 
$\delta \tau_H^0 \rightarrow \infty$ as 
$\eta /s \rightarrow \infty$. For practical purposes, 
it is therefore convenient to introduce a 
free-streaming regulator in the form of a threshold
value $\sigma_{\rm th}$. In this approximation, we may
consider instead that hydrodynamization is achieved 
when $\sigma$ drops below $\sigma_{\rm th}$ and the
corresponding time is denoted 
$\delta \tau_H^{\sigma_{\rm th}}$. In the case when 
$\sigma_{\rm th} < \sigma_{\rm min}$, we will take 
$\delta \tau_H^{\sigma_{\rm th} < \sigma_{\rm min}} 
= \delta \tau_H^0$, i.e. we will consider that
hydrodynamization is reached when 
$\sigma = \sigma_{\rm min}$. In the following, 
we will often employ $\sigma_{\rm th} = 0.01$, which 
is safely above the value of $\sigma_{\rm min}$
indicated by the blue points in 
Fig.~\ref{fig:1D_sigma}(a) for all values of 
$\eta / s$. However, for $\sigma_{\rm th} = 10^{-4}$,
Fig.~\ref{fig:1D_sigma} indicates that there will be
values of $\eta / s$ where 
$\delta \tau^{0.0001}_H = \delta \tau^{0}_H$.

As will be discussed in Sec.~\ref{sec:dtauH:res}, 
we assume that $\delta \tau_H^{\sigma_{\rm th}} \equiv 
\delta \tau_H^{\sigma_{\rm th}}(\tilde{w}_0^{-1})$ is 
a function only of the (inverse of the) initial value 
$\tilde{w}_0$ of the conformal variable \cite{Kamata:2020mka}
\begin{equation} 
 \tilde{w} = \frac{\tau T}{4 \pi \eta /s},
 \label{eq:wtilde_def}
\end{equation}
where $\pi \simeq 3.14$ should not be confused with the 
pressure anisotropy. 
In the perfect (inviscid) fluid limit, when $\eta /s = 0$,
hydrodynamization is instantaneous since
the pressure anisotropy satisfies $\pi = 0$ for all 
$\tau > \tau_0$. This gives the limit 
$\delta \tau^{\sigma_{\rm th}}_H(0) = 0$, regardless 
of the value of $\sigma_{\rm th}$. 
Away from $\tilde{w}_0^{-1} = 0$, 
$\eta / s$ can be considered as fixed, while 
$\tau_0 T_0$ are taken as large quantities, such that 
$\tilde{w}_0^{-1}$ remains small but finite. In this regime,
it is possible to estimate the hydrodynamization 
time $\delta \tau_H^{\sigma_{\rm th}}(\tilde{w}_0^{-1})$ 
based on a hydrodynamics transseries similar to the one 
derived in Ref.~\cite{Heller:2015dha}, as discussed in 
Sec.~\ref{sec:dtauH:trans}.
At the other end 
of the rarefaction spectrum, in the free streaming
limit, we have 
$\lim_{\tilde{w}_0^{-1} \rightarrow \infty} 
\delta \tau_H^0(\tilde{w}_0^{-1}) = \infty$, 
since the fluid cannot exhibit any attractor-like
behavior. Nevertheless, 
$\delta \tau_H^{\sigma_{\rm th}}$ takes finite values
when $\sigma_{\rm th}$ is kept finite. The values 
$\delta \tau_H^{\sigma_{\rm th}}$ will represent thus
maximum hydrodynamization times, which can be computed
exactly since the free-streaming limit can be obtained
analytically, as discussed in Secs~\ref{sec:dtauH:FS_hydro}
and \ref{sec:dtauH:FS_RTA} for the case of hydrodynamics and 
kinetic theory, respectively.

\subsection{Hydrodynamic limit: Transseries approach}
\label{sec:dtauH:trans}

In this section, we discuss the properties of 
$\delta \tau_H^{\sigma_{\rm th}}(\tilde{w}_0^{-1})$ at 
small values of $\tilde{w}_0^{-1}$. 
For the purpose of this section, we simplify the analysis 
by considering a conformally invariant system at vanishing 
chemical potential, such that $\tau_R = 5 (\eta /s) / T$.
In this regime, we can expect that second order hydrodynamics 
given by Eqs.~\eqref{eq:1D_cons} and \eqref{eq:1D_Pi} provides 
an adequate description. Taking the derivative of 
$\chi = (p + \pi)/ (p - \frac{\pi}{2})$ with respect to 
$\tau$, we obtain
\begin{equation}
 \frac{\tau \tilde{w}^{-1}}{4\pi} \frac{d\chi}{d\tau} = \frac{(1 - \chi)(2 + \chi)}{15} 
 - \frac{3\tilde{w}^{-1}}{70\pi} \left(1 + \frac{23 \chi}{3} + 
 \frac{2\chi^2}{3}\right),
 \label{eq:dchi_dtau}
\end{equation}
where $\pi \simeq 3.14$ appearing above should not be confused with 
the pressure anisotropy. Taking into account the relation 
\begin{equation}
 \tau \frac{d\tilde{w}}{d\tau} 
 = \frac{\tilde{w}(3 + \chi)}{2(2+\chi)},
 \label{eq:dwdtau}
\end{equation}
it can be seen that $\chi$ is a function only 
of $\tilde{w}$ by changing the derivative with respect 
to $\tau$ into a derivative with respect to $\tilde{w}$ 
in Eq.~\eqref{eq:dchi_dtau},
\begin{multline}
 \frac{3 + \chi}{8\pi} \frac{d \chi}{d\tilde{w}} = 
 \frac{(1 - \chi)(2 + \chi)^2}{15} \\ - 
 \frac{2 +\chi}{4\pi \tilde{w}} \frac{6}{35} \left(1 + 
 \frac{23 \chi}{3} + \frac{2\chi^2}{3}\right).
 \label{eq:dchi_dw}
\end{multline}
The large $\tilde{w}$ series solution of 
Eq.~\eqref{eq:dchi_dw},
\begin{equation}
 \chi(\tilde{w}) = 1 - \frac{2}{\pi \tilde{w}} + 
 \frac{6}{7 \pi^2 \tilde{w}^2}
 + O(\tilde{w}^{-3}),
 \label{eq:chi_w}
\end{equation}
is independent of the initial conditions and 
can be expected to have vanishing radius of convergence.

As argued in Ref.~\cite{Heller:2015dha}, $\chi(\tilde{w})$ 
can be more suitably represented as a transseries 
of the form
\begin{align}
 \chi(\tilde{w}) =& \sum_{m =0}^{\infty} c^m 
 \Omega^m(\tilde{w}) X_m(\tilde{w}), \nonumber\\
 X_m(\tilde{w}) =& \sum_{n = 0}^\infty 
 X_{m,n} \tilde{w}^{-n},\label{eq:chi_trans}
\end{align}
where $c$ is a constant related to the initial 
condition $\chi_0 \equiv \chi(\tilde{w}_0)$,
while $X_{m,n}$ are constants which are fixed 
order by order by the differential equation 
\eqref{eq:dchi_dw}. The function 
$\Omega(\tilde{w})$ controls the exponential
damping of deviations from the attractor solution 
and can be shown by direct substitution to satisfy
\begin{equation}
 \Omega(\tilde{w}) = \tilde{w}^{-\gamma}
 e^{-\tilde{w} \xi_0}, \qquad 
 \gamma = \frac{18}{35}, \qquad 
 \xi_0 = \frac{6\pi}{5}.
\end{equation}
The above result for $\xi_0$ is consistent 
with the one derived in Eq.~(11) of Ref.~\cite{Heller:2015dha} 
when using $C_{\tau\Pi} = 5 \eta /s$, while the difference in the 
exponent $\gamma$ can be explained by the discrepancy between the 
coefficient $\lambda = 38/21$ appearing in Eq.~\eqref{eq:1D_Pi} and 
the coefficient $4/3$ appearing in a similar term in Eq.~(4) of 
Ref.~\cite{Heller:2015dha}.
The $m = 0$ term in Eq.~\eqref{eq:chi_trans} is given by 
the series solution \eqref{eq:chi_w}, from where 
the coefficients $X_{m=0,n}$ can be easily read:
\begin{equation}
 X_{0,0} = 1, \qquad 
 X_{0,1} = -\frac{2}{\pi}, \qquad 
 X_{0,2} = \frac{6}{7 \pi^2}.
\end{equation}
At $m = 1$, there is an ambiguity in determining the 
leading order coefficient $X_{1,0}$, which can 
be resolved by essentially absorbing its value into the 
constant $c$ and setting $X_{1,0} = 1$. All other 
coefficients $X_{m,n}$ are then fixed by the 
differential equation \eqref{eq:dchi_dw}, e.g.:
\begin{align}
 X_{1,0} =& 1, & 
 X_{1,1} =& -\frac{3}{10 \pi}, & 
 X_{1,2} =& \frac{2657}{12600 \pi^2},\nonumber\\
 X_{2,0} =& \frac{5}{12}, & 
 X_{2,1} =& -\frac{5}{24 \pi}, & 
 X_{2,2} =& \frac{349}{1080 \pi^2}.
\end{align}

A more complex analysis based on the Borel transform 
and Pad\'e approximants presented in Ref.~\cite{Heller:2015dha}
is not necessary, since we are concerned with the properties 
of $\chi$ only at large initial values $\tilde{w}_0$ of the 
conformal parameter. In this regime, the (formally divergent) 
asymptotic series 
$X_m(\tilde{w})$ can be truncated at zeroth order, since the 
higher order terms represent corrections in 
powers of $\tilde{w}_0^{-1}$. The damping in the function 
$\Omega(\tilde{w})$ can in principle be offset by the constant 
$c$, which we relabel as
\begin{equation}
 c = \frac{\overline{c}}{\Omega(\tilde{w}_0)}.
\end{equation}
The ratio $\Omega(\tilde{w}) / \Omega(\tilde{w}_0)$ can be written as
\begin{equation}
 \frac{\Omega(\tilde{w})}{\Omega(\tilde{w}_0)} 
 = \left(\frac{\tau T}{\tau_0 T_0}\right)^{-\gamma} 
 \exp\left[-\xi_0 \tilde{w}_0\left(
 \frac{\tau T}{\tau_0 T_0} - 1\right)\right].
 \label{eq:Omega_ratio}
\end{equation}
Considering now $T = T_0 (\tau_0 / \tau)^{\frac{1}{3} - \delta}$, 
$\delta$ can be estimated from Eq.~\eqref{eq:dwdtau} via
\begin{equation}
 \delta \simeq \frac{\tau}{T} \frac{dT}{d\tau} + \frac{1}{3} = 
 \frac{1 - \chi}{6(2 + \chi)}.
\end{equation}
Close to the attractor solution, $\chi$ can be approximated 
by Eq.~\eqref{eq:chi_w} such that 
$\delta \simeq 1 / 9\pi \tilde{w}$, which becomes negligible when 
$\tilde{w}$ is large. Therefore, we consider as an approximation 
that $T / T_0 \simeq (\tau_0 / \tau)^{1/3}$ and estimate 
\begin{equation}
 \frac{\tau T}{\tau_0 T_0} \simeq 1 + \frac{2}{3} \delta \tau,
\end{equation}
where $\delta \tau = (\tau - \tau_0) / \tau_0$. 
At leading order, Eq.~\eqref{eq:Omega_ratio} simplifies to
\begin{equation}
 \frac{\Omega(\tilde{w})}{\Omega(\tilde{w}_0)} = 
 \exp\left(-\frac{2\xi_0}{3} \tilde{w}_0 \delta \tau\right).
\end{equation}
This suggests that, as $\tilde{w}_0$ increases, 
the product $\delta \tau_H^{\sigma_{\rm th}} \tilde{w}_0$ 
remains finite. Taking just the $m = 0$ and $m = 1$ terms in 
Eq.~\eqref{eq:chi_trans}, we have
\begin{equation}
 \chi(\tilde{w}) = X_0(\tilde{w}) + \overline{c}\,e^{-\frac{2\xi_0}{3} \tilde{w}_0 \delta \tau} X_1(\tilde{w}).\label{eq:chi_m1}
\end{equation}
Imposing $\chi = \chi_0$ when $\delta \tau = 0$, 
we find 
\begin{equation}
 \overline{c} = \frac{\chi_0 - X_0(\tilde{w}_0)}{X_1(\tilde{w}_0)}.
\end{equation}
This allows the standard deviation of $\chi$ to be expressed as
\begin{equation}
 \sigma(\chi) = \sigma(\chi_0) 
 e^{-\frac{2\xi_0}{3} \tilde{w}_0 \delta \tau} \frac{X_1(\tilde{w})}{X_1(\tilde{w}_0)},
\end{equation}
where $\sigma(\chi_0) = 1 / \sqrt{12}$ by direct computation. 
Imposing now $\sigma(\chi) = \sigma_{\rm th}$ and ignoring 
$\tilde{w}_0^{-1}$ corrections, the hydrodynamization time 
can be obtained as
\begin{equation}
 \delta\tau_H^{\sigma_{\rm th}} = \frac{5 \eta / s}{\tau_0 T_0} 
 \ln \left[\frac{\sigma(\chi_0)}{\sigma_{\rm th}}\right].
 \label{eq:dtauH_largew}
\end{equation}
The above equation shows that for any finite threshold 
$\sigma_{\rm th}$, $\delta \tau_H^{\sigma_{\rm th}}$ becomes 
proportional to $(4\pi \tilde{w}_0)^{-1} = \eta/s / (\tau_0 T_0)$
and reaches $0$ as $\tilde{w}_0^{-1} \rightarrow 0$. The apparent 
divergence of $\delta \tau_H^{\sigma_{\rm th}}$ as 
$\sigma_{\rm th} \rightarrow 0$ can be understood by noting that 
our ansatz in Eq.~\eqref{eq:chi_m1} assumes a smooth 
exponential decay toward the attractor for all initial 
conditions, which cannot account for the crossing and overshooting 
seen in Figures~\ref{fig:1D_profiles} and \ref{fig:1D_sigma}.

\subsection{Hydro: Free streaming limit}
\label{sec:dtauH:FS_hydro}

The free streaming (FS) limit can be obtained in the
framework of hydrodynamics by taking 
$\tau_R \rightarrow \infty$. This leaves 
Eq.~\eqref{eq:1D_cons} unchanged, while 
\eqref{eq:1D_Pi} becomes
\begin{equation}
 \tau \frac{\partial \pi}{\partial \tau} + 
 \frac{16 p}{15} + \frac{38 \pi}{21} = 0.
 \label{eq:FS_hydro_eq}
\end{equation}
The FS solution can be easily obtained as
\begin{align}
 p =& \frac{p_0}{(\tau / \tau_0)^{11/7}} 
 \left[\alpha \left(\frac{\tau}{\tau_0}\right)^\gamma 
 + (1 - \alpha) 
 \left(\frac{\tau}{\tau_0}\right)^{-\gamma}
 \right], \nonumber\\
 \pi =& \frac{p_0}{(\tau / \tau_0)^{11/7}} 
 \left[\alpha\left(\frac{5}{7} - 3\gamma\right)
 \left(\frac{\tau}{\tau_0}\right)^\gamma 
 \right.\nonumber\\
 &\left. + (1 - \alpha) \left(\frac{5}{7} 
 + 3\gamma\right)
 \left(\frac{\tau}{\tau_0}\right)^{-\gamma}\right],
 \label{eq:FS_hydro_sol}
\end{align}
where the exponent 
$\gamma = \sqrt{101} / 7 \sqrt{5} \simeq 0.642$ 
and the integration constant $\alpha$ is related 
to the initial anisotropic pressure $\pi_0$ via
\begin{equation}
 \alpha = \frac{1}{6\gamma}\left(\frac{5}{7} 
 + 3\gamma - \frac{\pi_0}{p_0}\right).
 \label{eq:FS_hydro_alpha}
\end{equation}
The time evolution of $\chi$ can be obtained by 
taking the ratio of $\mathcal{P}_L$ and 
$\mathcal{P}_T$ defined in Eq.~\eqref{eq:PLPT}. 
At large times, we find
\begin{equation}
 \chi = \chi_\infty + 
 \left(\frac{\tau_0}{\tau}\right)^{2\gamma}
 \Delta_\infty + 
 \left(\frac{\tau_0}{\tau}\right)^{4\gamma} 
 c_{4\gamma} + O(\tau^{-6\gamma}),
 \label{eq:FS_hydro_chi_asymp}
\end{equation}
where
\begin{align}
 \chi_\infty =& 2\frac{4- 7\gamma}{3 + 7\gamma} 
 \simeq -0.132, \nonumber\\
 \Delta_\infty =& \frac{1 - \alpha}{\alpha} 
 \frac{196 \gamma}{(3+7\gamma)^2}
 \simeq 2.241 \frac{1 - \alpha}{\alpha},
 \label{eq:FS_hydro_asymp2}
\end{align}
while the coefficient 
$c_{4\gamma} \equiv c_{4\gamma}(\alpha)$ 
is left unspecified.
It can be seen that in the FS limit, $\chi$ 
approaches a finite, negative value, instead of $0$
predicted by kinetic theory (discussed below). The 
value $\chi_\infty \simeq -0.132$ given above is
compatible with the limit 
$\mathcal{L}_1 / \mathcal{L}_0 = 
(\chi - 1) / (\chi + 2) \rightarrow -0.606$ derived 
in Ref.~\cite{Blaizot:2021}. The approach to this 
value is governed by a power law decay of exponent
$-2\gamma$. The information about the initial 
conditions is contained in the coefficient 
$\Delta_\infty$ of this transient term and hence 
is lost as $\tau \rightarrow \infty$. At large values 
of $\tau$, the standard deviation 
$\sigma(\chi) = [\braket{\chi^2} - 
\braket{\chi}^2]^{1/2}$ can be computed by noting that 
\begin{align}
 \braket{\chi^2} =& \chi_\infty^2 + 
 2\chi_\infty \braket{\Delta_\infty} 
 \left(\frac{\tau_0}{\tau}\right)^{2\gamma} \nonumber\\
 & + (2 \chi_\infty \braket{c_{4\gamma}} +
 \braket{\Delta_\infty^2})
 \left(\frac{\tau_0}{\tau}\right)^{4\gamma} 
 + O(\tau^{-6\gamma}),\nonumber\\
 \braket{\chi}^2 =& \chi_\infty^2 + 
 2\chi_\infty \braket{\Delta_\infty} 
 \left(\frac{\tau_0}{\tau}\right)^{2\gamma} \nonumber\\
 & + (2 \chi_\infty \braket{c_{4\gamma}} 
 + \braket{\Delta_\infty}^2) 
 \left(\frac{\tau_0}{\tau}\right)^{4\gamma} 
 + O(\tau^{-6\gamma}).\nonumber
\end{align}
Subtracting the above relations, we obtain:
\begin{equation}
 \sigma(\chi) = \sigma(\Delta_\infty)
 \left(\frac{\tau_0}{\tau}\right)^{2\gamma} 
 + O(\tau^{-3\gamma}),
 \label{eq:FS_hydro_sigma_asymp}
\end{equation}
where $\sigma(\Delta_\infty) = 
[\braket{\Delta_\infty^2} - 
\braket{\Delta_\infty}^2]^{1/2} \simeq 0.2591$, since 
$\braket{\Delta_\infty} \simeq 0.5920$ and 
$\braket{\Delta_\infty^2} \simeq 0.4176$. The
hydrodynamization timescale $\delta \tau_H^\infty$ 
for the FS regime of the second order hydrodynamics
theory can therefore be estimated for sufficiently 
small values of $\sigma_{\rm th}$ as
\begin{equation}
 \delta \tau_H^{\sigma_{\rm th}}(\infty) \simeq
 \left(\frac{\sigma(\Delta_\infty)}
 {\sigma_{\rm th}}\right)^{1/2\gamma} - 1.
 \label{eq:FS_hydro_dtauH_asymp}
\end{equation}

The hydrodynamization timescale 
$\delta\tau_H^{\sigma_{\rm th}}(\infty)$ can be found
for any value of $\sigma_{\rm th}$ by writing 
$\chi = (p + \pi) / (p - \pi/2)$ as a function of 
$\chi_0$ and $\tau$, using the exact solutions for 
$p$ and $\pi$ given in Eq.~\eqref{eq:FS_hydro_sol}.
Performing the $\chi_0$ integral numerically, 
$\sigma(\chi) \equiv \sigma(\chi; \tau)$ can be 
regarded as a function of $\tau$ and the 
hydrodynamization time 
$\delta \tau_H^{\sigma_{\rm th}}(\infty) =
(\tau^{\sigma_{\rm th}}_H - \tau_0) /\tau_0$ 
can be found using a numerical root finding algorithm
for the problem 
$\sigma(\chi; \tau^{\sigma_{\rm th}}_H) = 
\sigma_{\rm th}$. We find, e.g.,
\begin{align}
 \sigma_{\rm th} &= 10^{-2},& 
 \delta \tau_H^{0.01}(\infty)  &= 11.6492,\nonumber\\
 \sigma_{\rm th} &= 10^{-3},&
 \delta \tau_H^{0.001}(\infty) &= 74.785, \nonumber\\
 \sigma_{\rm th} &= 10^{-4},&
 \delta \tau_H^{0.0001}(\infty) &= 454.199,
 \label{eq:FS_hydro_dtauH}
\end{align}
in very good agreement with
Eq.~\eqref{eq:FS_hydro_dtauH_asymp}.
Because $\delta \tau_H^{\sigma_{\rm th}}(\infty)$ 
stays finite when $\sigma_{\rm th} > 0$, it is
reasonable to interpret $\sigma_{\rm th}$ as a FS
regulator.

\subsection{RTA: Free streaming limit}\label{sec:dtauH:FS_RTA}

In the case of the RTA, the exact solution of the
Boltzmann equation in the free streaming (FS) limit 
is \cite{Ambrus:2019wfc}
\begin{equation}
 f_{\rm FS} = \frac{g e^{\alpha_0}}{(2\pi)^3} 
 \exp \left(-\frac{k}{\Lambda_0} 
 \sqrt{1 + \zeta \xi^2}\right),
 \label{eq:FS_bjork}
\end{equation}
where $\zeta = \frac{\tau^2}{\tau_0^2}(1 + \xi_0) - 1$.
The longitudinal and transverse pressures can be 
derived analytically,
\begin{align}
 \mathcal{P}_L =& \frac{3g \Lambda_0^4 e^{\alpha_0}}
 {2\pi^2\zeta} 
 \left(\frac{\arctan\sqrt{\zeta}}{\sqrt{\zeta}} - 
 \frac{1}{1+\zeta}\right),\nonumber\\
 \mathcal{P}_T =& \frac{3g \Lambda_0^4 e^{\alpha_0}}
 {4\pi^2 \zeta} \left[1 + (\zeta - 1) 
 \frac{\arctan\sqrt{\zeta}}{\sqrt{\zeta}}\right],
 \label{eq:FS_RTA_sol}
\end{align}
while their ratio $\chi$ can be shown to obey 
\begin{equation}
 \chi = \frac{2}{1 + \xi_0} 
 \left(\frac{\tau_0}{\tau}\right)^2 - 
 \frac{8}{\pi(1 + \xi_0)^{3/2}} 
 \left(\frac{\tau_0}{\tau}\right)^{3} + O(\tau^{-4}).
\end{equation}
It is clear that $\chi \rightarrow 0$ as 
$\tau \rightarrow \infty$ and the transient term 
drops to $0$ faster than in the case of the hydro
solution in Eq.~\eqref{eq:FS_hydro_chi_asymp} 
($2\gamma \simeq 1.28$ compared to $2$ in the case 
of RTA). The leading term of $\sigma(\chi)$ is 
therefore given by 
\begin{equation}
 \sigma(\chi) = 2 \sigma\left(\frac{1}{1 + \xi_0}\right)
 \left(\frac{\tau_0}{\tau}\right)^2 \simeq 
 0.6871 \left(\frac{\tau_0}{\tau}\right)^2,
\end{equation}
where the integration with respect to $\chi_0$ was
performed by switching the integration variable in
Eq.~\eqref{eq:sigma_def} to $\xi_0$:
\begin{align}
 \braket{f(\xi_0)} =& \int_0^1 \d\chi_0\, 
 f(\xi_0) \nonumber\\
 =& -f(\xi_0= 0) - \int_0^\infty \d\xi_0\, 
 \chi_0 f'(\xi_0).
\end{align}
Using Eq.~\eqref{eq:RS_aux2} to express $\chi_0$ as 
a function of $\xi_0$, we find 
$\braket{(1 + \xi_0)^{-1}} = 0.1547$ and 
$\braket{(1 + \xi_0)^{-2}} = 0.1420$. The
hydrodynamization timescale 
$\delta \tau_H^{\sigma_{\rm th}}(\infty)$ can thus 
be estimated based on
\begin{equation}
 \delta \tau_H^{\sigma_{\rm th}}(\infty) \simeq
 \sqrt{\frac{2 \sigma[(1 +\xi_0)^{-1}]}
 {\sigma_{\rm th}}} - 1 \simeq 
 \frac{0.7666}{\sqrt{\sigma_{\rm th}}} - 1.
 \label{eq:FS_RTA_dtauH_asymp}
\end{equation}
Solving numerically $\sigma(\chi) = \sigma_{\rm th}$
starting from the exact solutions for $\mathcal{P}_L$
and $\mathcal{P}_T$ given in Eq.~\eqref{eq:FS_RTA_sol},
the following results can be obtained:
\begin{align}
 \sigma_{\rm th} &= 10^{-2},& 
 \delta \tau_H^{0.01}(\infty) &= 6.07422,\nonumber\\
 \sigma_{\rm th} &= 10^{-3},&
 \delta \tau_H^{0.001}(\infty) &= 22.6203, \nonumber\\
 \sigma_{\rm th} &= 10^{-4},&
 \delta \tau_H^{0.0001}(\infty) &= 75.0314.
 \label{eq:FS_RTA_dtauH}
\end{align}
The above values are in good agreement with those
obtained from Eq.~\eqref{eq:FS_RTA_dtauH_asymp}. 
In comparison to the results \eqref{eq:FS_hydro_dtauH}
obtained from the hydrodynamic equations, the values 
of $\delta \tau_H^{\sigma_{\rm th}}(\infty)$ obtained
from kinetic theory are notably smaller. It is
remarkable that the hydrodynamization time 
corresponding to the threshold 
$\sigma_{\rm th} = 10^{-2}$ remains extremely 
short even in the FS regime.

\subsection{Transition regime and scaling}
\label{sec:dtauH:res}

\begin{figure}
\begin{tabular}{c}
\includegraphics[width=0.95\columnwidth]{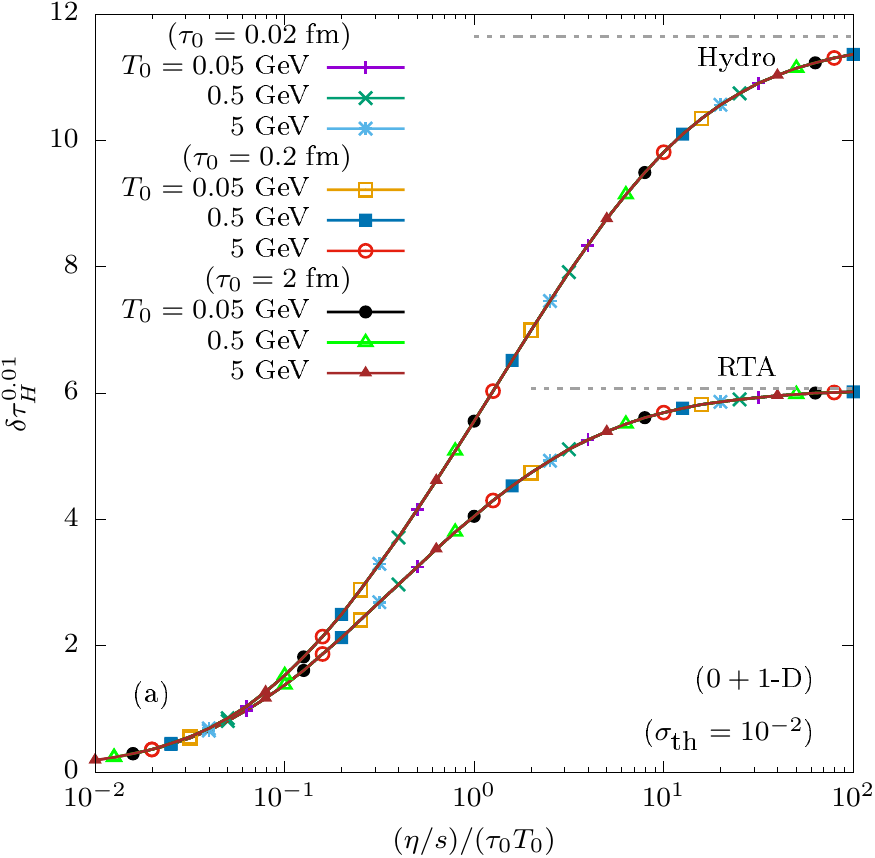} \\
\includegraphics[width=0.95\columnwidth]{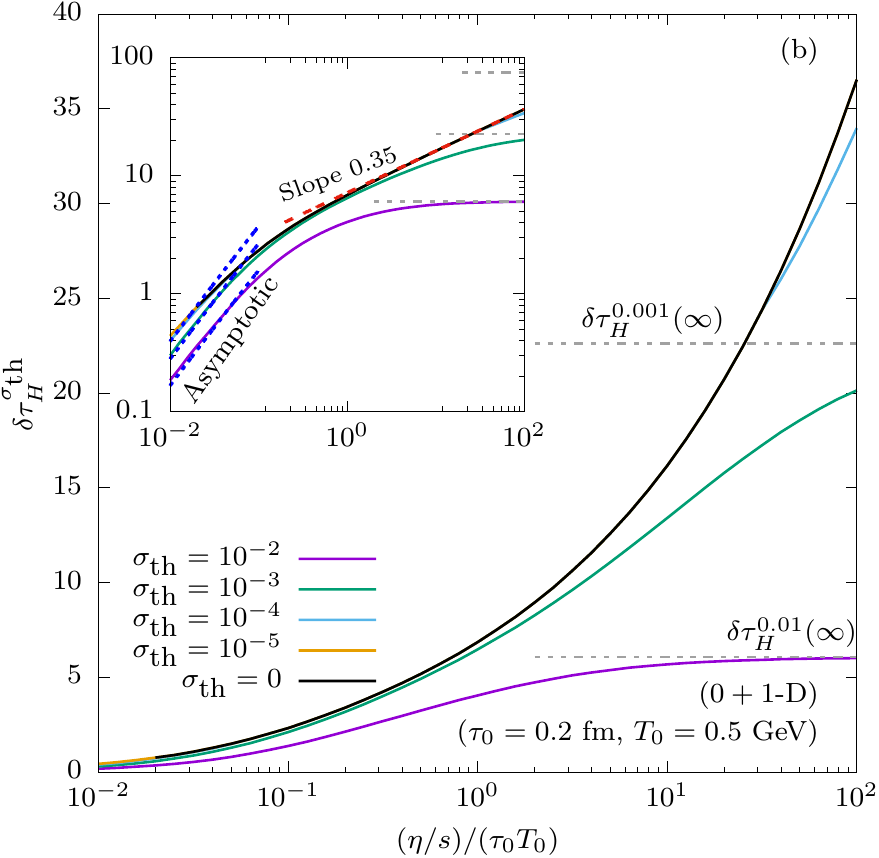}
\end{tabular}
\caption{
Dependence of $\delta \tau^{\sigma_{\rm th}}_H$ 
on $(\eta / s) / (\tau_0 T_0)$, (a) measured within
hydro (upper curves) and RTA (lower curves), 
for various values of $\tau_0$ and $T_0$ at 
$\sigma_{\rm th} =0.01$, with the horizontal dashed 
gray lines indicating the free-streaming limit 
$\delta\tau_H^{0.01}(\infty)$ given 
in Eqs.~\eqref{eq:FS_hydro_dtauH} and
\eqref{eq:FS_RTA_dtauH} for Hydro and RTA, 
respectively; 
(b) measured within RTA at $\tau_0 = 0.2\ {\rm fm}$ 
and $T_0 = 0.5\ {\rm GeV}$ for various values of 
$\sigma_{\rm th}$. The solid black line corresponds 
to the value $\delta \tau^0_H$ given by the condition 
$\sigma = \sigma_{\rm min}$, while the dashed horizontal 
gray lines represent the RTA free-streaming limits 
$\delta \tau_H^{0.01}$, $\delta \tau_H^{0.001}$ and $\delta \tau_H^{0.0001}$,
given in Eq.~\eqref{eq:FS_RTA_dtauH}. 
The inset shows the same 
plot in log-log scale, highlighting the
asymptotic limit for small $(\eta /s) / (\tau_0 T_0)$ 
derived in Eq.~\eqref{eq:dtauH_largew} with 
dotted blue lines. The dashed 
red line represents a polynomial fit to the $\delta\tau_H^0$ 
line for large values of $(\eta /s) / (\tau_0 T_0)$.}
\label{fig:1D_tauH}
\end{figure}

The analysis in the preceding subsection revealed that
the two limits, 
$\delta \tau_H^{\sigma_{\rm th}}(\tilde{w}_0^{-1} \rightarrow 0) = 0$ 
and $\delta \tau_H^{\sigma_{\rm th}}(\infty)$, are valid 
at any initial temperature $T_0$ or initial time 
$\tau_0$. At finite but small values of 
$\tilde{w}_0^{-1}$, Eq.~\eqref{eq:dtauH_largew} indicates 
that $\delta \tau_H^{\sigma_{\rm th}}$ is a function only 
of $(4\pi\tilde{w}_0)^{-1} = (\eta /s) / (\tau_0 T_0)$. 
This scaling is confirmed for both the hydrodynamics
equations (\ref{eq:1D_cons}, \ref{eq:1D_Pi}) and for 
the RTA in Fig.~\ref{fig:1D_tauH}(a). In this figure, 
we considered a $3 \times 3 = 9$ series of simulations
corresponding to initial times 
$\tau_0 \in \{0.02\ {\rm fm}$, $0.2\ {\rm fm}$, 
$2\ {\rm fm}\}$ and temperatures 
$T_0 \in \{0.05\ {\rm GeV}$, $0.5\ {\rm GeV}$, 
$5\ {\rm GeV}\}$. The ratio $\eta / s$ is taken 
such that the horizontal axis covers the range 
$10^{-2} \le \frac{\eta / s}{\tau_0 T_0} \le 10^2$. 
It can be seen that all curves are overlapped when
expressed with respect to $(\eta / s) / (\tau_0 T_0)$.

Next, we consider the dependence of 
$\delta \tau^{\sigma_{\rm th}}_H$ on the threshold 
$\sigma_{\rm th}$ below which hydrodynamization is
considered to be achieved. 
Figure~\ref{fig:1D_tauH}(b) shows that, as 
$\sigma_{\rm th}$ is decreased, 
$\delta \tau^{\sigma_{\rm th}}_H$ generally exhibits 
an increasing trend. This trend is stopped at the 
values of $\eta / s$ where 
$\sigma_{\rm th} < \sigma_{\rm min}$. This occurs at
intermediate values of $\eta / s$ first and extends
toward smaller and larger values of $\eta / s$ as 
$\sigma_{\rm th}$ is decreased, in agreement with the
qualitative picture painted by 
Fig.~\ref{fig:1D_sigma}(a). While 
$\delta \tau_H^{0.01}$ represents a good 
approximation for $\delta \tau_H^0$ only at very 
small values of $\tilde{w}_0^{-1}$, 
$\delta \tau_H^{0.001}$ gives similar values as 
$\delta \tau_H^0$ up to $(\eta /s ) / (\tau_0 T_0) \lesssim 1$,
while $\delta \tau_H^{0.0001}$ deviates from 
$\delta \tau_H^0$ only for $(\eta /s ) / (\tau_0 T_0) \gtrsim 20$. 
It is worth remarking that $\delta \tau_H^{0.01}$
reaches its asymptotic FS value for 
$(\eta /s ) / (\tau_0 T_0) \gtrsim 10$, while at 
$(\eta /s ) / (\tau_0 T_0) = 100$, $\delta \tau_H^{0.001}$ and 
$\delta \tau_H^{0.0001}$ are at $90\%$ and $45\%$ 
of their FS limits, respectively. 
The inset confirms that $\delta\tau_H^{\sigma_{\rm th}}$ scales 
linearly with $\tilde{w}_0^{-1}$ at small values of 
$\tilde{w}_0^{-1}$, as predicted by Eq.~\eqref{eq:dtauH_largew}.
Furthermore, we remark that at large $\tilde{w}_0^{-1}$,
$\delta \tau^{\sigma_{\rm th} = 0}_H$ seems to exhibit 
a polynomial growth 
$\delta \tau_H^0 \sim [(\eta / s) / 
(\tau_0 T_0)]^\alpha$ with $\alpha \simeq 0.35$, 
as indicated by the red dashed line.

Before ending this section, we discuss the procedure
employed to compute $\sigma(\chi)$. For each value of 
$\tau_0$, $T_0$ and $\eta / s$, a series of 
$1 \le i \le N_\chi$ simulations are performed, in 
which $\chi_0$ is initialized with the value 
$\chi_{0,i} = \frac{1}{N_\chi}(i - \frac{1}{2})$. 
These values are chosen to allow the integration with
respect to $\chi_0$ necessary for the computation of 
$\sigma(\chi)$ to be performed using the rectangle
method. In practice, we found that the maximum 
relative difference between the values of 
$\delta \tau_H$ computed based on $N_\chi = 5$ and $10$
intervals was below $1\%$ for RTA and below $1.2\%$ 
for hydro. The results shown in Fig.~\ref{fig:1D_tauH}
are for definiteness computed with $N_\chi = 10$
intervals.

\section{Bjorken flow with transverse expansion}\label{sec:transv}

\begin{figure}[!t]
\begin{tabular}{c}
\includegraphics[width=0.95\columnwidth]{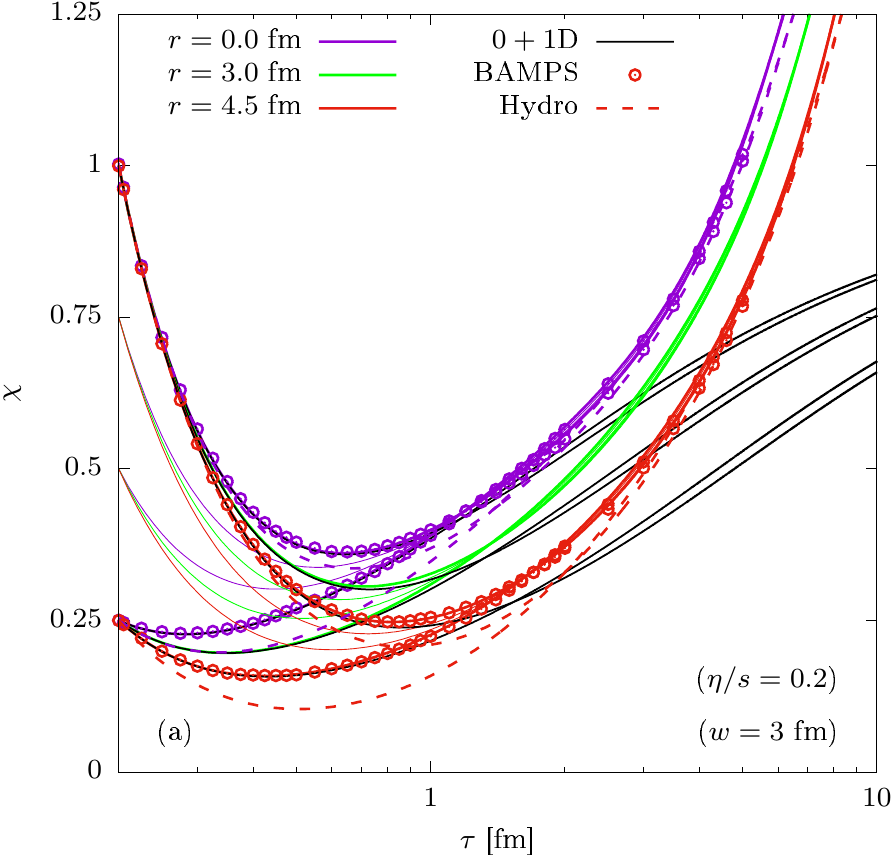} \\
\includegraphics[width=0.95\columnwidth]{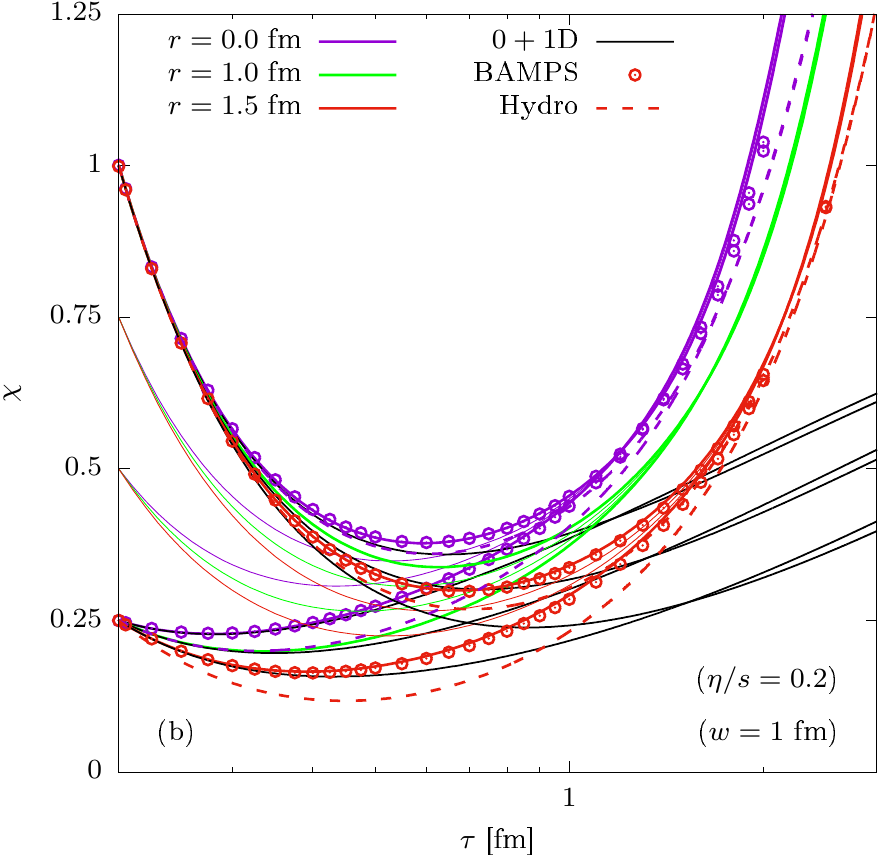}
\end{tabular}
\caption{
The ratio $\chi = \mathcal{P}_L / \mathcal{P}_T$ for 
(a) $w = 3\ {\rm fm}$ and (b) $w = 1\ {\rm fm}$,
obtained using the RTA (solid lines), BAMPS 
(red circles) and hydro (dashed lines), calculated at 
$r = 0$ (purple), $r=w$ (green), and $r = 3w/2$ (red),
and represented with respect to the Bjorken time $\tau$.
The RTA solution for the corresponding $0+1{\rm D}$
system is shown using the solid black lines.
\label{fig:2D_profiles}
}
\end{figure}

In Sec.~\ref{sec:bjork}, we considered
``hydrodynamization'' (or memory-loss with respect 
to the initial pressure anisotropy $\chi_0$) due 
solely to the longitudinal expansions. In this section,
we consider the same problem in a system undergoing 
also transverse expansion, by initializing a
longitudinally boost-invariant system with 
transverse Gaussian density and temperature profiles:
\begin{equation}
 n_0(r) = n_0(0) e^{-r^2 / w^2}, \qquad
 T_0(r) = T_0(0) e^{-r^2 / 3w^2}, 
 \label{eq:gw_T0}
\end{equation}
The initial time is set for definiteness to 
$\tau_0 = 0.2\ {\rm fm}$ and we consider that the 
system is homogeneous with respect to the rapidity. 
The width parameter $w$ is set to $3\ {\rm fm}$ and 
$1\ {\rm fm}$, corresponding roughly to 
${\rm Au}+{\rm Au}$ and $p+p$ collisions, 
respectively \cite{Gallmeister:2018mcn}.

We first consider $\eta/s=0.2$, in order to be close 
to the favored values describing the ''fluid'' behavior
observed in ultra-relativistic heavy ion collisions
\cite{Romatschke:2007mq,Xu:2007jv,Schenke:2010rr,Uphoff:2014cba}. 
In Fig.~\ref{fig:2D_profiles}, we show the typical
hydrodynamization dynamics occurring at various 
distances from the origin, namely $r = 0$, $w$ and 
$3w/2$. The fate of the fluid at larger values of 
$r$ is less important, since the disks within 
$r = w$ and $3w/2$ contain $74\%$ and $95\%$,
respectively, of the total energy available in 
the transverse plane.
Because of the $r$-dependence of the initial state, 
the local conditions in each of these points are
different. 

The evolution of $\chi$ can be divided into three parts.
In the initial stage, corresponding to small values of 
$\tau$, the system dynamics is dictated by the
longitudinal expansion, following closely the 
$0+1{\rm D}$ results. After an intermediate stage, 
$\chi$ increases significantly faster than in the
$0+1{\rm D}$ case, signaling that the system dynamics 
is then dominated by the transverse expansion of the
fireball. Values of $\chi$ larger than 1 can be seen,
since $\mathcal{P}_T$ is depleted at a faster rate 
than $\mathcal{P}_L$ as the transverse dynamics 
become dominant. As in the $0+1{\rm D}$ case shown 
in Fig.~\ref{fig:1D_profiles}, the RTA and BAMPS 
results are in excellent agreement. Even though the
hydro results show some discrepancy during the
longitudinal expansion-dominated phase, they agree 
with both BAMPS and RTA data at large values of $\tau$.
We remark that even at $r = 3w/2$, the RTA and BAMPS
curves corresponding to $w = 3\ {\rm fm}$ 
[Fig.~\ref{fig:2D_profiles}(a)] follow closely the
$0+1{\rm D}$ curves all the way until hydrodynamization.
By contrast, in the $w= 1\ {\rm fm}$ scenario, there is
a clear departure between the curves corresponding to
the simulation with transverse dynamics and the 
$0+1{\rm D}$ case. Still, hydrodynamization can be 
seen to take place on a similar timescale.

\begin{figure*}[!t]
\begin{center}
\begin{tabular}{cc}
\includegraphics[width=0.45\linewidth]{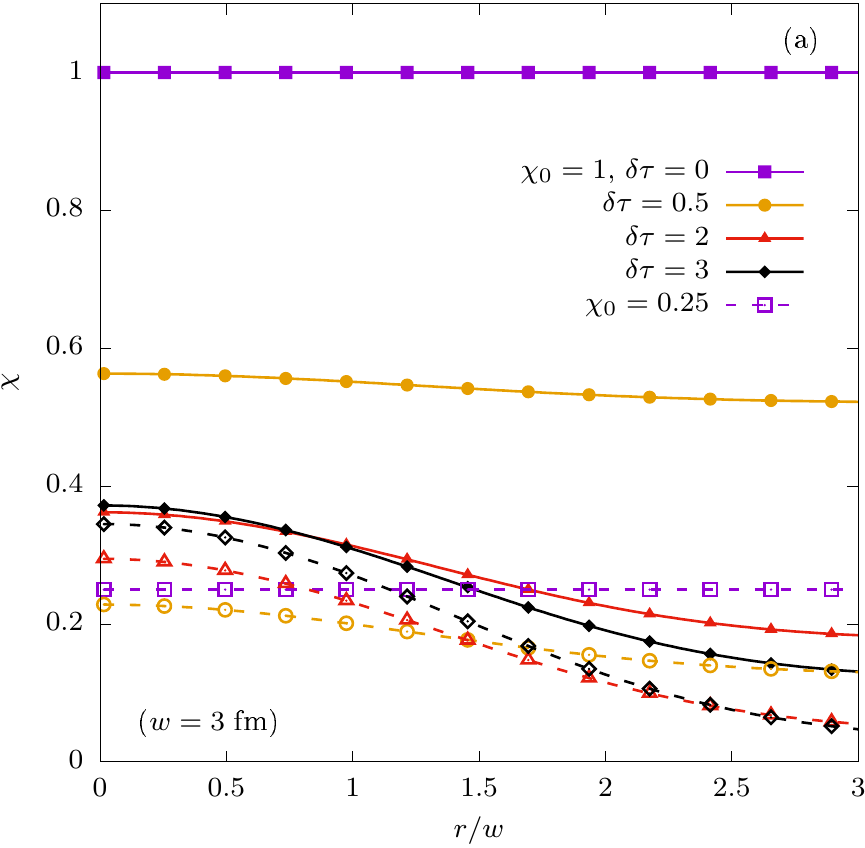} & 
\includegraphics[width=0.45\linewidth]{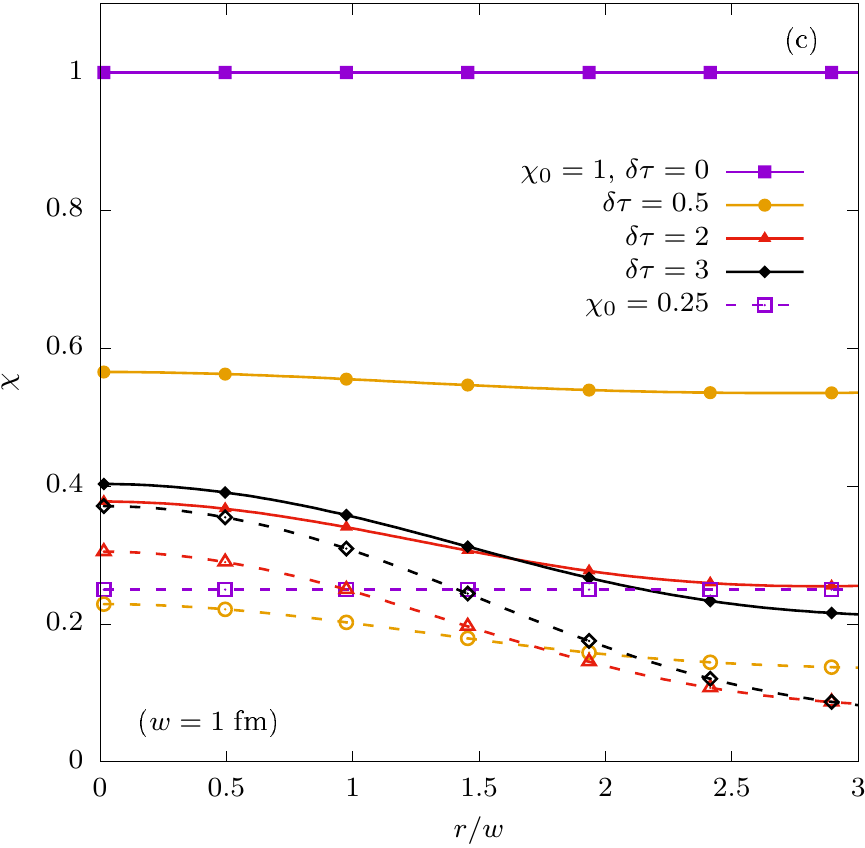}\\
\includegraphics[width=0.45\linewidth]{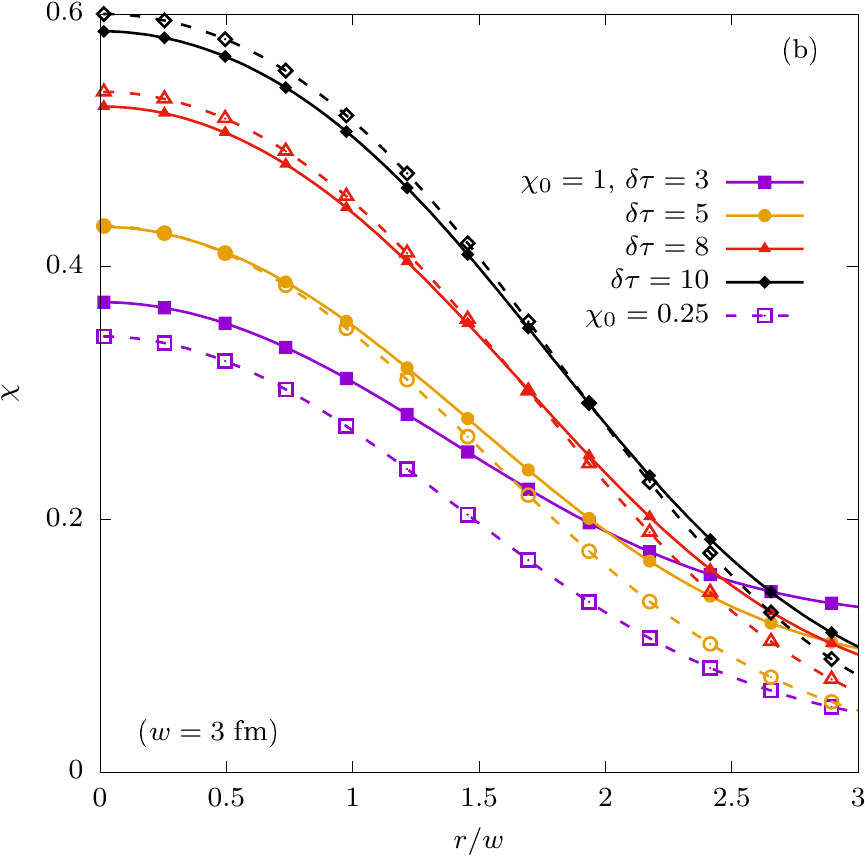} & 
\includegraphics[width=0.45\linewidth]{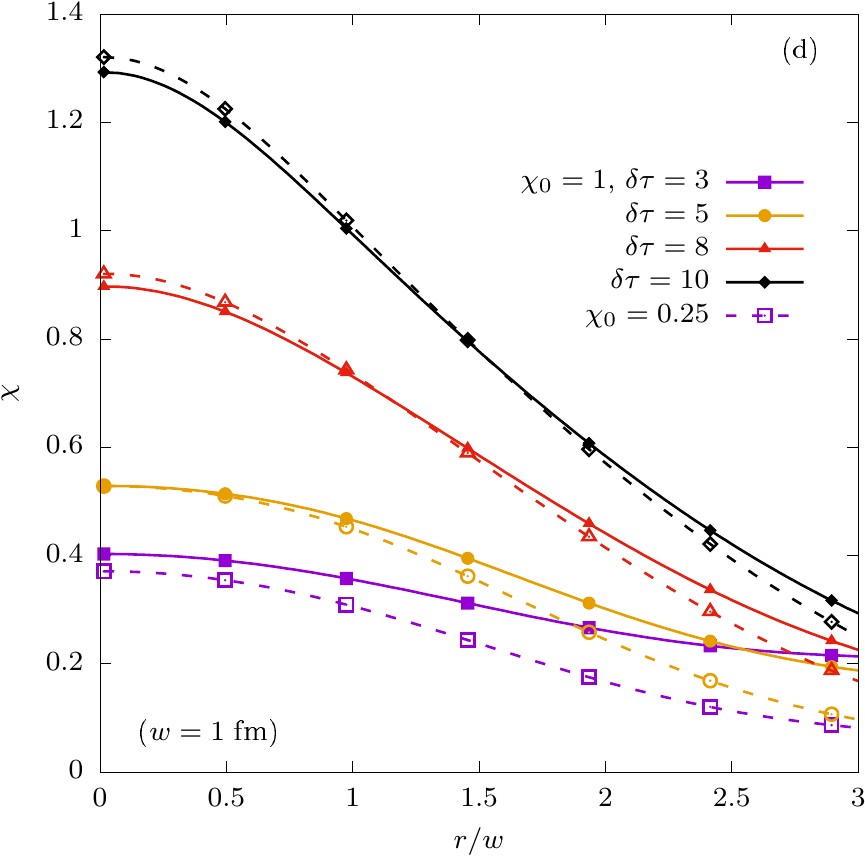}\\
\end{tabular}
\end{center}
\caption{Radial profiles of $\chi$ at different values
of $\delta \tau = (\tau - \tau_0) / \tau_0$ for 
$w = 3\ {\rm fm}$ (left column) and $1\ {\rm fm}$ 
(right column). The initial conditions are 
$\tau_0 = 0.2\ {\rm fm}$ and $T_0 = 0.5\ {\rm GeV}$,
while $\chi_0 = 1$ for the solid curves with filled
symbols and $\chi_0 = 0.25$ for the dotted curves 
with empty symbols. The ratio $\eta / s$ is $0.2$.
\label{fig:radial}}
\end{figure*}

Note however that the hydrodynamization occurs at
different times for different radii: hydrodynamization
in the central region starts earlier than in the
intermediate region, and the latest in the outmost
region, as we will discuss in more detail below.

As an additional remark for the situation of both 
the small and the large systems, but more 
significantly for the smaller system, in the outer
regions the hydrodynamization occurs at stages when 
the energy density has already dropped below values of 
$1\,{\rm GeV/fm}^3$. This behavior might challenge 
the hydrodynamical simulations of $p+p$ or $p+A$
collisions.

In order to gain more insight on the radial 
dependence of $\chi$, Fig.~\ref{fig:radial} shows 
the radial profiles of $\chi$ at various 
values of $\tau$, corresponding to the initial
conditions $\chi_0 = 1$ (solid lines) and 
$\chi_0 = 0.25$ (dashed lines). 
The left and right columns show the 
$w = 3\ {\rm fm}$ and $w = 1\ {\rm fm}$ systems,
respectively. The top line [panels (a) and (c)]
represent the early time evolution of $\chi$. 
It can be seen that at small times, the evolution 
of $\chi$ is similar between the $w = 3$ and 
$1\ {\rm fm}$ simulations. At 
$\delta \tau = (\tau - \tau_0) / \tau_0 = 3$, it can be
seen that the $\chi_0 = 1$ and $\chi_0 = 0.25$ curves
are very close to each other around $r = 0$. However,
the distance between these curves increases with $r$,
indicating that hydrodynamization is more rapid at the
fireball center than at the system periphery. On the
lower line of Fig.~\ref{fig:radial}, the same
hydrodynamization can be seen to be achieved at
increasingly large $r$ as $\delta \tau$ is increased.
This is in line with the analysis of the $0+1{\rm D}$ system
from Sec.~\ref{sec:bjork}, since the hydrodynamization
time $\delta \tau_H$ is expected to increase due to the
increase of the local value of 
$\tilde{w}_0^{-1}(r) = (4\pi \eta / s) / [\tau_0 T_0(r)]$. 
A key difference between the larger ($w = 3\ {\rm fm}$)
and smaller ($w = 1\ {\rm fm}$) systems is that 
$\chi$ increases much faster in the latter case. 
This is because the transverse expansion is driven by
larger gradients, becoming dominant compared to the
longitudinal expansion at a faster rate than in the
larger system.

\begin{figure}[!t]
\begin{center}
\begin{tabular}{c}
\includegraphics[width=0.9\linewidth]{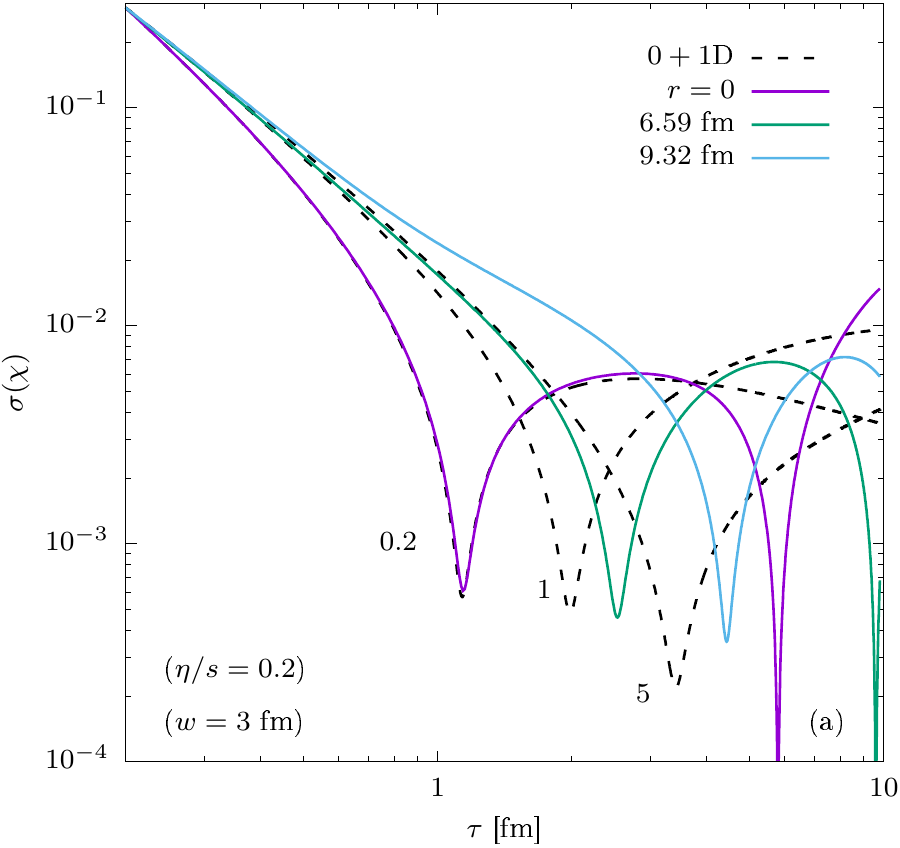} \\
\includegraphics[width=0.9\linewidth]{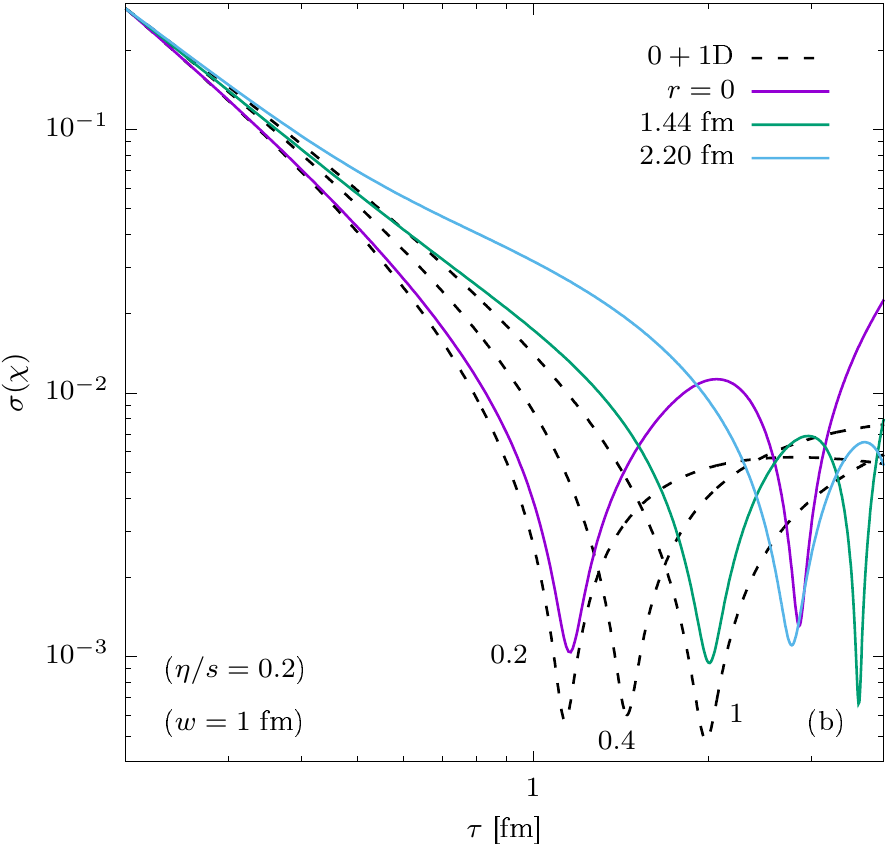}
\end{tabular}
\end{center}
\caption{Evolution of $\sigma(\chi)$ measured at various
values of $r$ for transversely expanding systems with 
$\eta / s = 0.2$, having widths (a) $w = 3\ {\rm fm}$
and (b) $w = 1\ {\rm fm}$. The values of $r$ are chosen
such that $\tilde{w}_0^{-1}(r) = (4\pi\eta /s) / 
[\tau_0 T_0(r)]$ matches that of the $0+1{\rm D}$ 
system with $T_0 = 0.5\ {\rm GeV}$ and the values of 
$\eta / s$ inscribed next to the $0+1{\rm D}$ lines
(shown with dotted black lines). The initial time is 
$\tau_0 = 0.2\ {\rm fm}$.
\label{fig:sigma_etas0p2}}
\end{figure}

Focusing now on the systems with $\eta / s = 0.2$, 
we investigate the evolution of $\sigma(\chi)$ at
various distances $r$ from the fireball center in
Fig.~\ref{fig:sigma_etas0p2}. In the $w = 3\ {\rm fm}$
system, shown in panel (a), $r = 0$ corresponds to 
the fireball center, while $r = 6.59\ {\rm fm}$ and 
$9.32\ {\rm fm}$ correspond to initial values of the
local temperature $T_0(r) = 0.1$ and $0.02\ {\rm GeV}$,
respectively. In the $w = 1\ {\rm fm}$ system, the
values $1.44\ {\rm fm}$ and $2.20\ {\rm fm}$ of $r$
correspond to $T_0(r) = 0.25\ {\rm GeV}$ and 
$0.1\ {\rm GeV}$. The black dotted lines represent
results obtained in the $0+1{\rm D}$ system, 
initialized such that the values of 
$\tilde{w}_0^{-1} = (4\pi\eta / s) / \tau_0 T_0$ match 
those of the points considered in the transversely
expanding systems. In particular, we kept 
$\tau_0 = 0.2\ {\rm fm}$ and $T_0 = 0.5\ {\rm GeV}$
fixed and considered $\eta /s = 0.2$, $1$ and $5$ for 
$w = 3\ {\rm fm}$, while the values $\eta / s= 0.2$,
$0.4$ and $1$ were employed for the $w = 1\ {\rm fm}$
system.

\begin{figure}[!t]
\begin{center}
\begin{tabular}{c}
\includegraphics[width=0.9\linewidth]{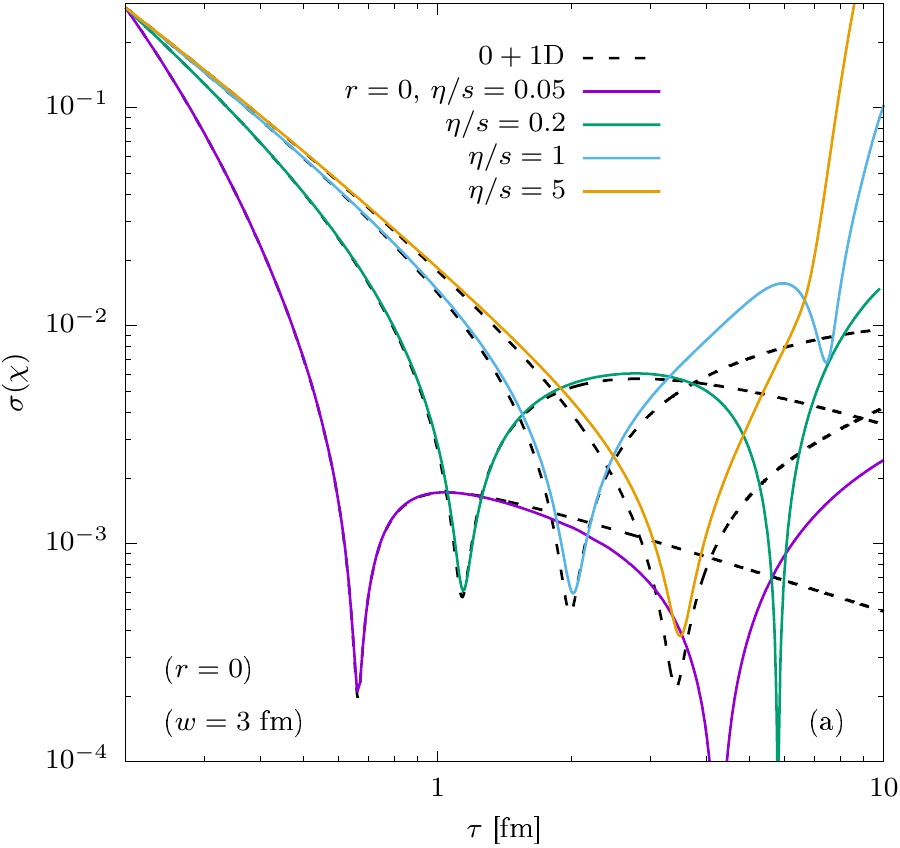} \\
\includegraphics[width=0.9\linewidth]{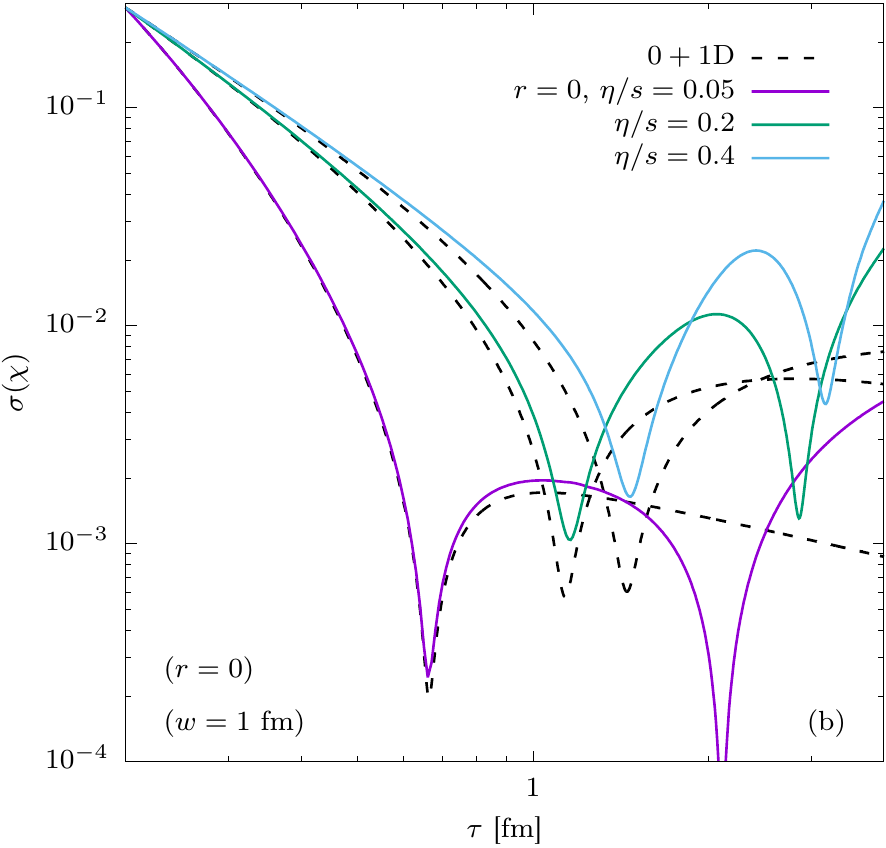}
\end{tabular}
\end{center}
\caption{Evolution of $\sigma(\chi)$ measured at 
$r= 0$ for transversely expanding systems with 
various values of $\eta / s$, having widths (a) 
$w = 3\ {\rm fm}$ and (b) $w = 1\ {\rm fm}$. The 
dotted lines represent results from the $0+1{\rm D}$
system having $T_0 =0.5\ {\rm GeV}$ and the same 
value of $\eta / s$ as that in the transversely
expanding system. The initial time is 
$\tau_0 = 0.2\ {\rm fm}$.
\label{fig:sigma_r0}}
\end{figure}

As in the $0+1{\rm D}$ system, $\sigma(\chi)$ exhibits 
a decrease toward a minimum value $\sigma_{\rm min}$
reached after a relatively short time. At $r = 0$, 
the approach to this minimum is almost identical in 
the $1+1{\rm D}$ system as in the $0+1{\rm D}$ system.
As $r$ is increased, the agreement deteriorates and 
the minimum is reached at a later time. The effect 
is more pronounced for the smaller system, where 
the transverse gradients are stronger, which indicates
that the effect of the transverse expansion is to 
delay hydrodynamization in comparison to the prediction
of the $0+1{\rm D}$ model.

A remarkable feature seen for the $r = 0$ curve in
panel (a) of Fig.~\ref{fig:sigma_etas0p2} is that
a second minimum emerges at later times, namely at 
$\tau = 5.78\ {\rm fm}$ and $2.50\ {\rm fm}$ for the 
$w = 3\ {\rm fm}$ and $1\ {\rm fm}$ systems,
respectively. From Fig.~\ref{fig:2D_profiles}, 
it can be seen that at these times, ${\bar \chi}$ is
around $1.18$ and $1.58$ for the larger and smaller
system, respectively, thus the system evolution at 
this stage is dominated by the transverse dynamics.
Thus, the second minima seen in 
Fig.~\ref{fig:sigma_etas0p2} reveals a new attractor
solution which is due to the transverse expansion of 
the system.

We now focus on the dynamics at the center of the
fireball and consider systems with various values of 
$\eta /s$. A comparison between the $1+1{\rm D}$ and 
the corresponding $0+1{\rm D}$ systems is presented 
in Fig.~\ref{fig:sigma_r0}. For the larger system 
($w = 3\ {\rm fm}$), shown in panel (a), good agreement
can be seen even at $\eta / s = 5$. In the smaller
system, a discrepancy can be seen at the level of 
the value of $\sigma_{\rm min}$, which increases at
larger $\eta / s$. However, for $\eta / s \lesssim 0.4$,
the hydrodynamization time 
$\delta \tau_H^{\sigma_{\rm th} = 0}$ (when $\sigma$
reaches the local minimum $\sigma_{\rm min}$) remains
similar to that of the $0+1{\rm D}$ system. We can thus
conclude that the approach to the attractor solution is
dominated for both large and small systems by the
longitudinal dynamics of the $0+1{\rm D}$ system. 
For the larger system, the analogy holds up to very 
high values of $\eta / s$, while the smaller system
exhibits more visible deviations even at small 
$\eta / s$. It is notable that the second minima 
emerges significantly faster in the smaller system 
than in the larger system, indicating that
hydrodynamization due to transverse expansion 
is more effective here.

\begin{figure}[!t]
\begin{tabular}{c}
\includegraphics[width=0.95\columnwidth]{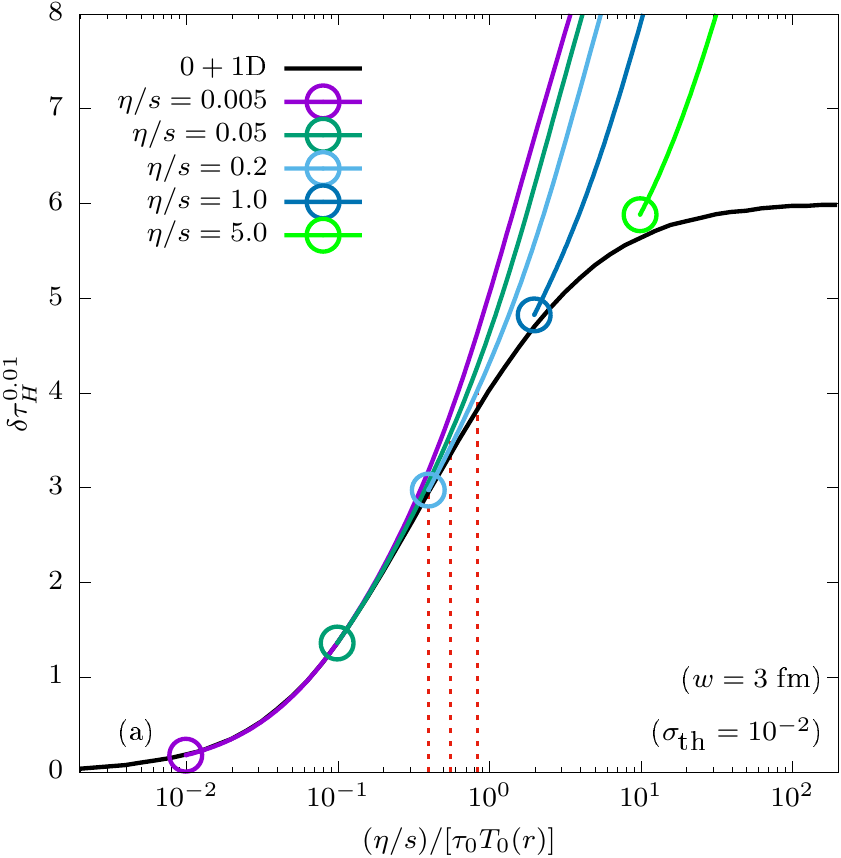} \\
\includegraphics[width=0.95\columnwidth]{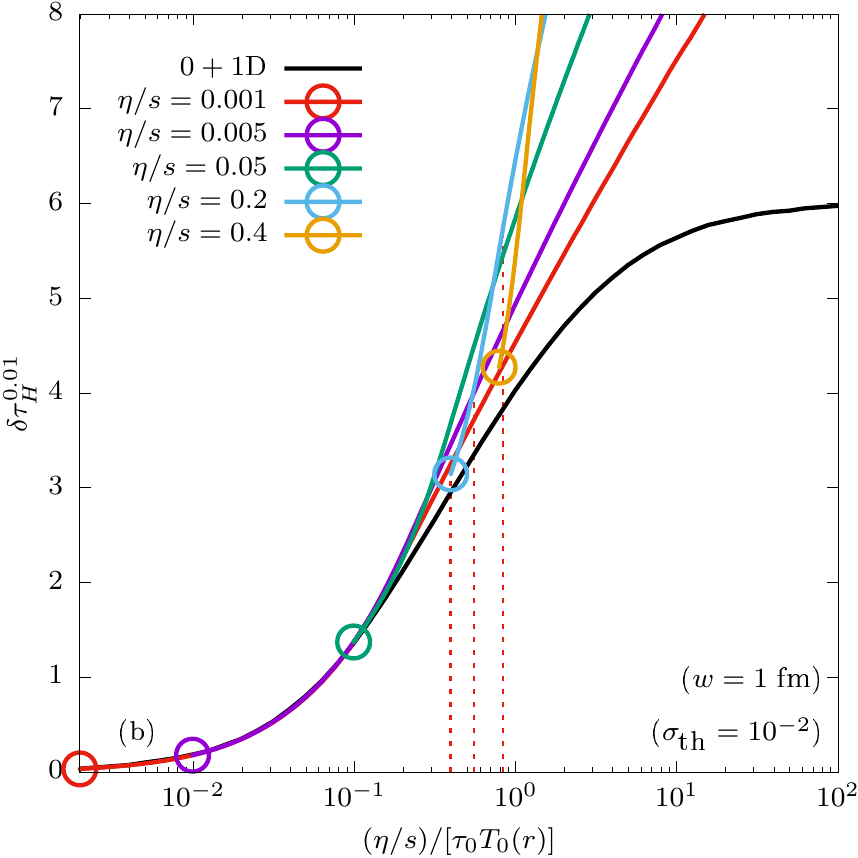}
\end{tabular}
\caption{
The hydrodynamization timescale $\delta \tau^{0.01}_H$
corresponding to a threshold $\sigma_{\rm th} = 0.01$
for the $w = 3\ {\rm fm}$ (a) and $w = 1\ {\rm fm}$ 
(b) systems, represented with respect to 
$(\eta / s) / [\tau_0 T_0(r)]$,
where the initial temperature $T_0(r)$ is given by
Eq.~\eqref{eq:gw_T0}. The circles represent 
$\delta \tau^{0.01}_H$ corresponding to the central
point ($r = 0$) of the fireball and each curve
corresponds to a different value of $\eta / s$. 
The red dashed lines mark the positions $r = 0$, $w$ 
and $3w/2$ for the system with $\eta / s= 0.2$,
corresponding to the curves shown in 
Fig.~\ref{fig:2D_profiles}.
\label{fig:2D_tauH}
}
\end{figure}

Figure~\ref{fig:2D_tauH} shows the hydrodynamization
timescale $\delta \tau^{0.01}_H$ (corresponding to 
$\sigma_{\rm th} = 0.01$) in the scenario with
transverse expansion achieved from the RTA approach 
as a function of $\tilde{w}_0^{-1}(r) = 
(4\pi\eta / s) / [\tau_0 T_0(r)]$. The results for the
larger ($w = 3\ {\rm fm}$) and smaller 
($w = 1\ {\rm fm}$) systems are shown in panels (a) 
and (b), respectively. The initial temperature 
$T_0(r)$ decreases with increasing $r$, as indicated 
in Eq.~\eqref{eq:gw_T0}. We considered simulations 
with $\eta /s$ between $0.001$ and $5$. For each value
of $\eta / s$, the simulation covers the $x$-axis range
from $\tilde{w}_0^{-1}(0) = (4\pi\eta / s) / 
[\tau_0 T_0(0)]$ up to infinity. It can be seen that 
$\delta \tau_H^{0.01}$ for the $1+1{\rm D}$ system 
is very similar to that for the $0+1{\rm D}$ system at
small values of $\tilde{w}_0^{-1}(r)$. 

For fixed $\eta / s$, larger deviations between the
$0+1{\rm D}$ and $1+1{\rm D}$ results appear as either
$r$ is increased or $w$ is decreased. When 
$\delta \tau_H^{0.01}$ for a given point in the 
system exceeds a certain threshold value 
($\delta \tau_H \gtrsim 3$ and $2$ for $w = 3$ and 
$1\ {\rm fm}$, respectively), a deviation with respect
to the $0+1{\rm D}$ results toward higher values of 
$\delta \tau_H$ can be seen. Furthermore, there are 
always points which are sufficiently far from the 
origin to exhibit deviations from the $0+1{\rm D}$
prediction in their hydrodynamization timescale. 

\begin{figure}[!t]
\begin{tabular}{c}
\includegraphics[width=0.95\columnwidth]{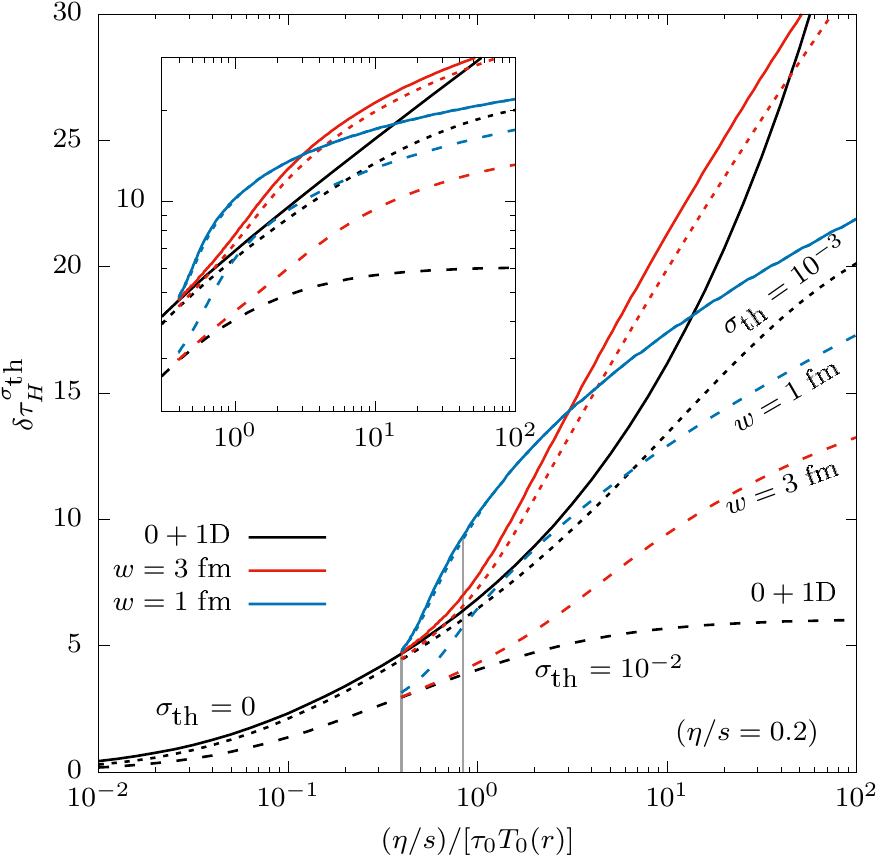} 
\end{tabular}
\caption{
Hydrodynamization time $\delta \tau_H^{\sigma_{\rm th}}$
with respect to $(\eta / s) / [\tau_0 T_0(r)]$
for the $w = 3\ {\rm fm}$ (red lines)
and $1\ {\rm fm}$ (blue lines) systems at 
$\eta / s = 0.2$. The results for 
$\sigma_{\rm th} = 0.01$, $0.001$ and $0$ 
(corresponding to $\sigma_{\rm min}$) are shown with
dashed, dotted and continuous lines, respectively. 
The black lines represent the $0+1{\rm D}$ results from
Fig.~\ref{fig:1D_tauH}(b). The gray lines delimit the
region between the fireball center and $r = 3w/2$. The
inset shows the same plot in log-log scale.
\label{fig:2D_tauH_sth}
}
\end{figure}

We now consider the dependence of 
$\delta \tau_H^{\sigma_{\rm th}}$ on the threshold 
value $\sigma_{\rm th}$, represented in 
Fig.~\ref{fig:2D_tauH_sth}. As already seen in
Fig.~\ref{fig:1D_tauH}(b), decreasing 
$\sigma_{\rm th}$ causes 
$\delta \tau_H^{\sigma_{\rm th}}$ to increase toward
the $\delta \tau_H^{\sigma_{\rm th} = 0}$ limit,
achieved when $\sigma_{\rm th} < \sigma_{\rm min}$ 
(more details regarding this notation are given 
in Sec.~\ref{sec:dtauH:def}). The analysis focuses on
the $\eta /s = 0.2$ system, for which $95\%$ of the 
initial fireball energy (contained within 
$r \lesssim 1.5 w$) is between 
$0.4 \lesssim (\eta/s)/[\tau_0 T_0(r)] \lesssim 0.84$,
indicated as gray lines in the figure. In this region,
it can be seen that $\delta \tau_H^{\sigma_{\rm th}}$
for the larger system behaves essentially as predicted
by the $0+1{\rm D}$ system. For the smaller system, 
$\delta \tau_H^{\sigma_{\rm th}}$ is close to the
$0+1{\rm D}$ prediction at the fireball center,
increasing to a value about $50\%$ larger than the
$0+1{\rm D}$ prediction at $r = 3w/2$.

For both the larger and the smaller systems, 
$\delta \tau_H^{\sigma_{\rm th}}$ at 
$\sigma_{\rm th} = 10^{-3}$ is almost equal to its 
limit value $\delta \tau_H^0$, being further from this
limit for the larger system than for the smaller 
system (for the latter, the curves corresponding to 
$\sigma_{\rm th} = 0.001$ and $0$ are almost
overlapped). This indicates that $\sigma_{\rm min}$ has
a larger value for the $w = 1\ {\rm fm}$ system compared
to that for $w  = 3\ {\rm fm}$, as also seen in
Figs.~\ref{fig:sigma_etas0p2} and \ref{fig:sigma_r0}.
Furthermore, while for $\sigma_{\rm th} = 0.01$, the 
$w = 1\ {\rm fm}$ value for 
$\delta \tau_H^{\sigma_{\rm th}}$ is larger than the
value corresponding to $w = 3\ {\rm fm}$ over the 
whole domain considered in Fig.~\ref{fig:2D_tauH_sth},
at $\sigma_{\rm th} = 0$ it can be seen
that $\delta \tau_H^0$ becomes smaller for
$w = 1\ {\rm fm}$ when $r \simeq 2.51 w$. The time
coordinate corresponding to this point, where 
$\delta \tau_H^0 \simeq 14.45$, is 
$\tau \simeq 3.09\ {\rm fm}$. As seen in 
Fig.~\ref{fig:2D_profiles}, for the smaller system, 
the transverse expansion is already dominant, which 
may explain why hydrodynamization is accelerated
compared to the larger system, which is still in 
a transition phase from longitudinally dominated 
to transversally dominated expansion.

\section{Conclusion} \label{sec:conc}

In this work, we considered the problem of
hydrodynamization in a system of a conformal 
ideal gas of ultrarelativistic particles undergoing
boost-invariant longitudinal expansion with and 
without transverse dynamics. Quantitatively, we
described hydrodynamization on the basis of a
(nondimensional) timescale 
$\delta \tau^{\sigma_{\rm th}}_H = 
(\tau_H^{\sigma_{\rm th}} - \tau_0) / \tau_0$, defined 
in terms of the time $\tau^{\sigma_{\rm th}}_H$ in which
the standard deviation $\sigma(\chi)$ of the ratio 
$\chi = \mathcal{P}_L / \mathcal{P}_T$ with respect to
its initial value ($0 \le \chi_0 \le 1$ were considered)
either reaches its minimum value $\sigma_{\rm min}$,
corresponding to the (imperfect) merger of this family
of curves, or decreases below a threshold value 
$\sigma_{\rm th}$. 

In the conformal limit of the $0+1{\rm D}$ problem, 
$\delta \tau^{\sigma_{\rm th}}_H$ is a function only 
of the conformal parameter 
$\tilde{w}_0^{-1} = (4\pi\eta / s) / (\tau_0 T_0)$. 
With respect to this parameter, 
$\delta \tau^{0.01}_H$ (obtained for 
$\sigma_{\rm th} = 0.01$) is bounded between 
two limits, $\delta \tau_H^{0.01}(0) = 0$ and 
$\delta \tau_H^{0.01}(\infty) \simeq 6$ 
(about $1.2\ {\rm fm}$ after initial time 
$\tau_0 = 0.2\ {\rm fm}$), corresponding to the 
inviscid and free-streaming regimes, respectively. 
For the system with transverse dynamics, there appears 
a competition between the $0+1{\rm D}$ hydrodynamization
timescale and the timescale associated with transverse
dynamics.

In the $1+1{\rm D}$ setup, we described the initial
transverse distribution of Gaussian form with widths 
$w = 1$ and $3\ {\rm fm}$, corresponding to small
($p+p$) and large ($A+A$) collisions. A comparison
between the results obtained with the three numerical
schemes considered in this paper (Hydro, RTA and BAMPS)
is presented in Fig.~\ref{fig:2D_profiles}. While Hydro
presents some deviations from RTA and BAMPS, the RTA
results follow closely the BAMPS results in all tested
flow regimes (see Figs.~\ref{fig:1D_profiles}, 
\ref{fig:2D_profiles}, and \ref{fig:1D_HS}). 
It is worth stressing that the excellent agreement
between RTA and BAMPS recommends RTA as a simulation 
tool for this type of systems, since it is a
significantly faster numerical method than BAMPS. 

For the points with sufficiently small values of 
$(\eta / s) / [\tau_0 T_0(r)]$ (where $T_0(r)$ is 
the local initial temperature), the hydrodynamization
time is very well approximated by the $0+1{\rm D}$
prediction. With increasing values of the radius, 
$\delta \tau_H^{\sigma_{\rm th}}$ deviates from the
$0+1{\rm D}$ prediction to larger values, faster for 
the smaller system than for the larger one. For the
system with larger transverse size ($w = 3\ {\rm fm}$),
we found that at $\eta / s = 0.2$, the hydrodynamization
of the region $r < 3w/2$ (containing $95\%$ of the
initial energy of the fireball) follows very closely 
the $0+1{\rm D}$ dynamics. While the center of the
smaller fireball is also well captured by the 
$0+1{\rm D}$ dynamics, hydrodynamization times of up 
to $50\%$ larger can be seen around $r = 3w/2$,
indicating that the transverse dynamics has the 
effect of slowing down hydrodynamization due to
longitudinal expansion. Our analysis revealed the
emergence of a second minimum of $\sigma(\chi)$,
suggesting the existence of an attractor due to 
the transverse expansion.

In our picture of a heavy-ion collision, our results
indicate that for a given $\eta/s$ one always finds a
radius in the overlap region beyond which the 
transversal dynamics become dominant and the
hydrodynamization is delayed compared to the 
innermost region of the fireball. For the 
outermost regions, this has to be confronted 
also with the timescales connected with the 
decrease of energy density and freeze out, making 
the situation challenging for a hydrodynamical
description in the case of very small systems.

\begin{acknowledgments}
V.E.A. gratefully acknowledges the support of the
Alexander von Humboldt Foundation through a Research
Fellowship for postdoctoral researchers. J.A.F., K.G., 
and C.G. acknowledge support by the Deutsche 
Forschungsgemeinschaft (DFG, German Research Foundation) 
through the CRC-TR 211 'Strong-interaction matter under 
extreme conditions'-- project number 315477589 -- TRR 211. 
J.A.F acknowledges support from the 'Helmholtz Graduate 
School for Heavy Ion research'. K.G. was supported by
the Bundesministerium f\"ur Bildung und Forschung
(BMBF), Grant No. 3313040033. This work was supported by
the Helmholtz Research Academy Hessen for FAIR (HFHF).
\end{acknowledgments}

\appendix

\section{$0+1{\rm D}$ BJORKEN FLOW FOR HARD-SPHERES}
\label{app:HS}

\begin{figure}
\begin{tabular}{c}
\includegraphics[width=0.95\columnwidth]{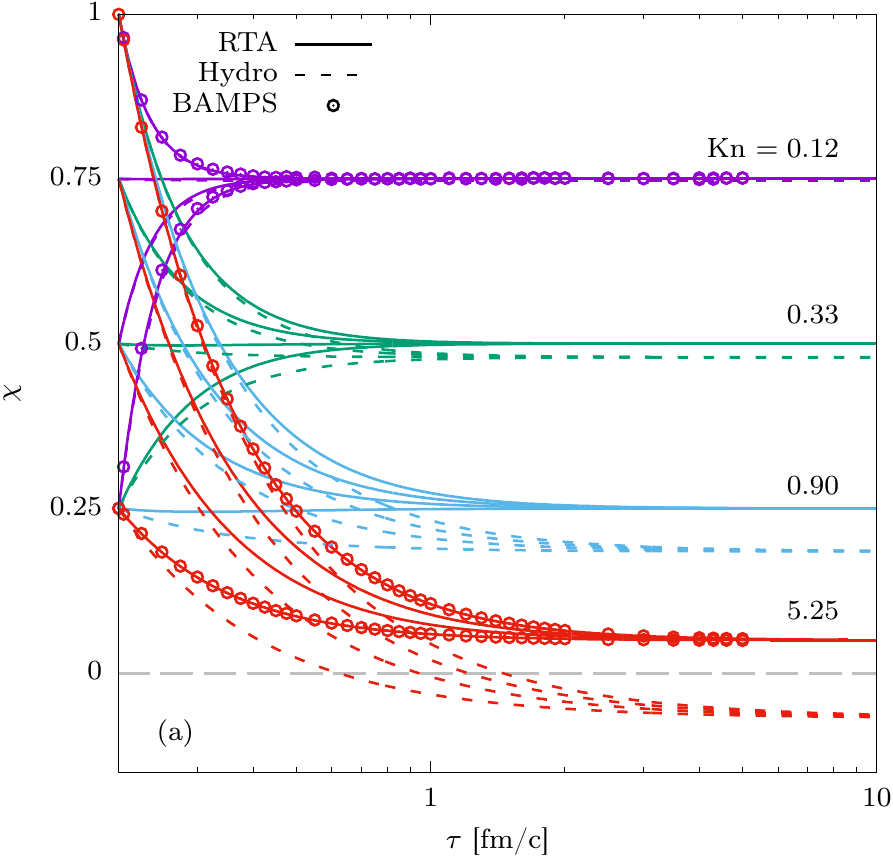}
\end{tabular}
\caption{
Evolution of the pressure anisotropy 
$\chi = \mathcal{P}_L / \mathcal{P}_T$ with respect 
to the Bjorken time $\tau$ in the context of the 
hard-sphere gas for various values of the Knudsen 
number ${\rm Kn}$ \eqref{eq:HS_Kn}. The initial
conditions are $\tau_0 = 0.2\ {\rm fm}$ and 
$T_0 = 0.5\ {\rm GeV}$. The RTA and hydro results 
are shown with solid and dashed lines, respectively,
while the BAMPS results are shown using empty circles.
\label{fig:1D_HS}
}
\end{figure}

\begin{figure}
\begin{tabular}{c}
\includegraphics[width=0.95\columnwidth]
{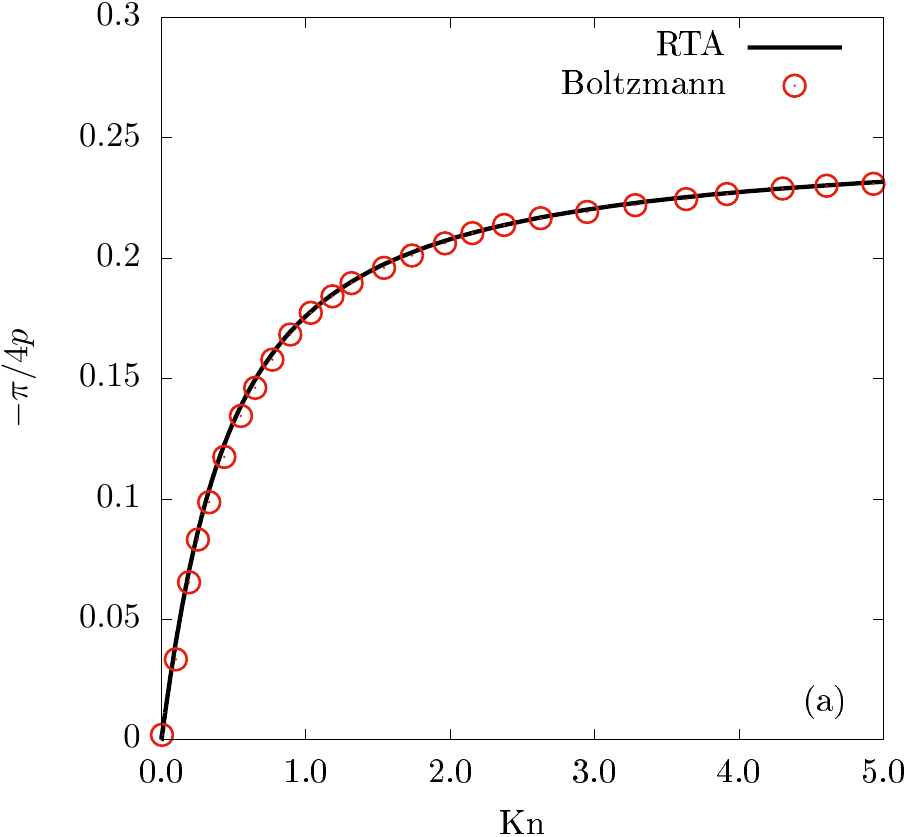}
\end{tabular}
\caption{
The asymptotic value of $-\pi / 4p = (1 - \chi) / 
[2(2 + \chi)]$ as a function of the Knudsen number 
${\rm Kn}$ \eqref{eq:HS_Kn}. Our results obtained 
using RTA are shown with the black line, while the
results reported by Denicol and Noronha 
\cite{Denicol:2019lio} are shown with red circles.
\label{fig:1D_vs_Denicol}
}
\end{figure}

In this section, we consider the $0+1{\rm D}$ Bjorken
flow of a gas of hard spheres (HS), as described, 
e.g., by Denicol and Noronha in 
Ref.~\cite{Denicol:2019lio}. In the context of the 
BAMPS approach, the collision cross section is set to 
a constant value $\sigma$. The degree of rarefaction 
can be conveniently characterized by the Knudsen number
${\rm Kn}$, defined as \cite{Denicol:2019lio}
\begin{equation}
 {\rm Kn} = \frac{1}{n \tau \sigma} = 
 \frac{1}{n_0 \tau_0 \sigma}.
 \label{eq:HS_Kn}
\end{equation}
In the Hydro setup, the HS gas can be implemented by
noting that the shear viscosity is related to 
$\sigma$ via \cite{Denicol:2012cn}
\begin{equation}
 \eta = \frac{1.2654 T}{\sigma},\label{eq:HS_eta}
\end{equation}
where $T = p / n = p \tau / n_0 \tau_0$ is the local
temperature. In the RTA approach, the HS gas is
simulated by setting the relaxation time $\tau_R$
according to Eq.~\eqref{eq:RTA_eta_tau}, with 
$\eta$ given by Eq.~\eqref{eq:HS_eta}.

In Fig.~\ref{fig:1D_HS}, the Hydro, RTA, and BAMPS
results for $\chi = \mathcal{P}_L / \mathcal{P}_T$ are
compared for various values of ${\rm Kn}$. The initial
conditions are set as in Fig.~\ref{fig:1D_profiles},
namely $\tau_0 = 0.2\ {\rm fm}$ and 
$T_0 = 0.5\ {\rm GeV}$. The values of ${\rm Kn}$ are
chosen such that the asymptotic value of $\chi$ is
$0.75$, $0.5$, $0.25$ and $0.05$. As also noted for the
case of the gas with constant $\eta / s$ reported in
Fig.~\ref{fig:1D_profiles}, all three methods agree at
small ${\rm Kn}$. The hydro results exhibit a departure
from the RTA and BAMPS results already at 
${\rm Kn} = 0.33$, achieving negative values for 
$\chi$ when ${\rm Kn} = 5.25$. The agreement between 
RTA and BAMPS remains excellent for both small and 
large values of ${\rm Kn}$. 

As remarked in Ref.~\cite{Denicol:2012cn} and confirmed
in Fig.~\ref{fig:1D_HS}, $\chi$ reaches a constant value
as $\tau \rightarrow \infty$, which depends on 
${\rm Kn}$. In Fig.~\ref{fig:1D_vs_Denicol}, we compare
our RTA results for 
$-\Pi / 4p = (1 - \chi) / [2(2 + \chi)]$  with the
results computed on the basis of the full Boltzmann
collision integral for the hard sphere gas in
Ref.~\cite{Denicol:2012cn}, finding excellent agreement
throughout the whole Knudsen range (between $0$ and
$5$).

\section{NUMERICAL METHOD FOR THE RTA}\label{app:RTA}

In this section, we present the details of the numerical
method employed to solve the relativistic Boltzmann
equation in the Anderson-Witting relaxation time
approximation \cite{Anderson:1974,Anderson:1974b}. 
The method is inspired by the finite difference Lattice
Boltzmann (LB) algorithm 
\cite{Romatschke:2011hm,Ambrus:2012high,Ambrus:2018kug,Succi:2018,Gabbana:2019ydb,Gabbana:2020oka}.

The strategy for devising the numerical method is split
into three main parts described in this appendix. The
derivation of the relativistic Boltzmann equation in the
context of the longitudinal boost-invariant system with
transverse expansion is presented in
Sec.~\ref{app:RTA:vielbein}. At the heart of this
derivation is the vielbein formalism 
\cite{Cardall:2002bp}, which allows spherical
coordinates to be employed in the momentum space
together with curvilinear spatial coordinates
\cite{Ambrus:2018kug}. 

The momentum space discretization is based on Gauss
quadratures for the integration with respect to
spherical coordinates and follows LB methodology
\cite{Ambrus:2012high,Romatschke:2011hm,Ambrus:2018kug},
being described in Sec.~\ref{app:RTA:discretization}.
The algorithm for computing the derivatives with respect
to the momentum space degrees of freedom, appearing
due to the use of a curvilinear coordinate system, 
is also discussed here.

The spatial and temporal discretization, as well as the
numerical schemes employed for the advection and time
stepping, are briefly summarized in
Sec.~\ref{app:RTA:numsch}.

\begin{table}
\begin{tabular}{ll||llll}
 $w$ (${\rm fm}$) \hspace{5pt} & $\eta / s$ \hspace{5pt} & $S$ & $\delta \tau / \tau_0$\hspace{5pt}  & $Q_\varphi$ \hspace{5pt} & $Q_\xi$ \hspace{5pt} \\\hline
 $1$ & $\le 0.005$ & $200$\hspace{5pt} & $0.005$ & $20$ & $80$ \\
 $1$ & $> 0.005$ & $200$ & $0.0025$ & $80$ & $160$ \\
 $3$ & (all values) & $200$ & $0.01$ & $40$ & $160$
\end{tabular}
\caption{Parameters used for the RTA simulations
presented in Sec.~\ref{sec:transv}. See
Appendix~\ref{app:RTA} for the 
intepretation of the above notation.
\label{tab:params}}
\end{table}

The parameters employed for the simulations discussed in
Sec.~\ref{sec:transv} are summarized for convenience in
Table~\ref{tab:params}. We tested that the simulation
results were within $1\%$ errors compared to the values
obtained by doubling the resolution in any of the
numerical parameters shown in Table~\ref{tab:params}.

\subsection{Separation of variables in momentum space
using the vielbein formalism}
\label{app:RTA:vielbein}

The relativistic Boltzmann equation can be written with
respect to the Minkowski (Cartesian) coordinates 
$(t, x, y, z)$ as follows:
\begin{equation}
 k^\mu \partial_\mu f = C[f],\label{eq:boltz}
\end{equation}
where $k^\mu = (k^t, k^x, k^y, k^z)$ represent the
Cartesian momentum-space components and $C[f]$ is 
the collision integral (discussed below). 

In a system with longitudinal boost invariance, it 
is convenient to employ the Bjorken coordinates in
Eq.~\eqref{eq:ds2}. Moreover, in this paper, we consider
systems with azimuthal symmetry in the transverse plane.
Thus, the macroscopic observables depend only on Bjorken
time $\tau$ and on the radial distance $r$. The line
element \eqref{eq:ds2} becomes
\begin{equation}
 \d s^2 = \d\tau^2 - \d r^2 - r^2 \d\theta^2 - 
 \tau^2 \d\eta_s^2.
\end{equation}
In order to take advantage of this symmetry in the 
full phase-space, the momentum space degrees of 
freedom can be chosen with respect to the following
vielbein field (tetrad),
\begin{align}
 e_\htau =& \partial_\tau, &
 e_\hatr =& \partial_r, &
 e_\htheta =& r^{-1} \partial_\theta, &
 e_{\heta_s} =& \tau^{-1} \partial_{\eta_s},\nonumber\\
 \omega^\htau =& d\tau, &
 \omega^\hatr =& dr, &
 \omega^\htheta =& r d\theta, &
 \omega^{\heta_s} =& \tau d\eta_s.
 \label{eq:vielbein}
\end{align}
The tetrad components 
$k^\halpha = k^\mu \omega^\halpha_\mu$ are then employed
to perform the momentum space integration, such that the
particle four-flow vector $N^\halpha$ and the 
stress-energy tensor $T^{\halpha\hbeta}$ are computed 
as follows:
\begin{align}
 N^\halpha =& \int \frac{\d^3k}{k^\htau} f\, 
 k^\halpha, &
 T^{\halpha\beta} =& \int \frac{\d^3k}{k^\htau} f\,
 k^\halpha k^\hbeta.
 \label{eq:macro_vielb}
\end{align}
The hatted indices are raised and lowered with the
Minkowski metric $\eta^{\halpha\hbeta} = 
{\rm diag}(1, -1, -1, -1)$, i.e. 
$N^\halpha = \eta^{\halpha\hbeta} N_\hbeta$.
In order to perform the integrals in
Eq.~\eqref{eq:macro_vielb}, it is convenient to
introduce spherical coordinates in the momentum 
space, via
\begin{equation}
 \begin{pmatrix}
  k^\hatr \\ k^\htheta
 \end{pmatrix} = k \sqrt{1 - \xi^2} 
 \begin{pmatrix}
  \cos\varphi \\ \sin\varphi
 \end{pmatrix}, \qquad 
 k^{\heta_s} = k \xi.
\end{equation}
In the case of the ultrarelativistic gas, $k^\htau = k$. 

We are now ready to write down the relativistic
Boltzmann equation for the distribution function
$f(x^\mu, k^{\widetilde{\imath}})$ with the 
phase-space dependence on the curvilinear coordinates
$x^\mu = (\tau, r, \theta, \eta_s)$ and the momentum
space degrees of freedom 
$k^{\widetilde{\imath}} = (k, \xi, \varphi)$. 
It is based on the general theory developed by Cardall 
and Mezzacappa \cite{Cardall:2002bp} and employed also
in Ref.~\cite{Ambrus:2018kug}:
\begin{multline}
 \frac{1}{\sqrt{-g}} 
 \partial_\mu (\sqrt{-g} e^\mu_\halpha k^\halpha f) - 
 \frac{k^\htau}{\sqrt{\lambda}} 
 \frac{\partial}{\partial k^{\widetilde{\imath}}}
 \left(K^{\widetilde{\imath}}{}_\hati 
 \Gamma^\hati{}_{\halpha\hbeta} 
 \frac{k^\halpha k^\hbeta}{k^\htau} 
 f \sqrt{\lambda}\right) \\
 = C[f],\label{eq:boltz_vielb}
\end{multline}
where 
$\lambda^{-1/2} = 
|{\rm det} K^{\widetilde{\jmath}}{}_\hati|$ and 
the matrix $K^{\widetilde{\jmath}}{}_{\hati} = 
\partial k^{\widetilde{\jmath}} / \partial k^\hati$,
computed in Eq.~(2.20) of Ref.~\cite{Ambrus:2018kug}, 
is reproduced below for convenience:
\begin{equation}
 K^{\widetilde{\jmath}}{}_{\hati} = 
 \begin{pmatrix}
  \cos\varphi \sqrt{1 - \xi^2} & 
  \sin\varphi \sqrt{1 - \xi^2} & 
  \xi \\
  -\frac{\xi}{k} \cos\varphi\sqrt{1 - \xi^2} & 
  -\frac{\xi}{k} \sin\varphi\sqrt{1 - \xi^2} & 
  \frac{1 - \xi^2}{k} \\
  -\frac{\sin\varphi}{k\sqrt{1 - \xi^2}} & \frac{\cos\varphi}{k\sqrt{1 - \xi^2}} & 
  0
 \end{pmatrix}.
 \label{eq:Pij}
\end{equation}
The connection coefficients 
$\Gamma^\hati{}_{\halpha\hbeta}$ appearing in
Eq.~\eqref{eq:boltz_vielb} can be computed via
\begin{equation}
 \Gamma^\hsigma{}_{\hbeta\hgamma} = 
 \frac{1}{2} \eta^{\hsigma\halpha}
 (c_{\halpha\hbeta\hgamma} + 
 c_{\halpha\hgamma\hbeta} - 
 c_{\hbeta\hgamma\halpha}), 
\end{equation}
where the Cartan coefficients are based on the
commutators of the vielbein tetrad vectors, 
$[e_\halpha, e_\hbeta]= c_{\halpha\hbeta}{}^\hgamma
e_\hgamma$. Based on Eq.~\eqref{eq:vielbein}, 
we find $c_{\hat{\tau} \hat{\eta}_s \hat{\eta}_s} 
= -c_{\hat{\eta}_s \hat{\tau} \hat{\eta}_s} = \tau^{-1}$ 
and $c_{\hat{r} \hat{\theta} \hat{\theta}} = 
-c_{\hat{\theta} \hat{r} \hat{\theta}} = r^{-1}$, with all 
other Cartan coefficients vanishing, leading to
\begin{equation}
 \Gamma^\htau{}_{\heta_s \heta_s} = 
 \Gamma^{\heta_s}{}_{\htau\heta_s} = \tau^{-1}, \qquad 
 \Gamma^\hatr{}_{\htheta\htheta} = 
 -\Gamma^\htheta{}_{\hatr\htheta} = -r^{-1}.
 \label{eq:Gamma}
\end{equation}
Plugging now Eqs.~\eqref{eq:Pij} and \eqref{eq:Gamma}
into Eq.~\eqref{eq:boltz_vielb}, we find
\begin{multline}
 \frac{1}{\tau} \frac{\partial (f\tau)}{\partial \tau} + 
 \frac{k^\hatr}{r k^\htau} 
 \frac{\partial (f r)}{\partial r} + 
 \frac{k^\htheta}{r k^\htau} 
 \frac{\partial f}{\partial \theta} + 
 \frac{k^{\heta_s}}{\tau k^\htau} 
 \frac{\partial f}{\partial \eta} \\
 - \frac{\xi^2}{\tau k^2} 
 \frac{\partial(f k^3)}{\partial k} - 
 \frac{1}{\tau} 
 \frac{\partial[\xi(1 - \xi^2) f]}{\partial \xi} -
 \frac{k\sqrt{1 - \xi^2}}{r k^\htau} 
 \frac{\partial(f \sin\varphi)}{\partial \varphi} \\
 = \frac{1}{k^\htau} C[f].
 \label{eq:boltz_gen}
\end{multline}
In the case of massless particles (considered 
throughout this paper and in what follows), 
$m = 0$ and $k^\htau = k$. 

The collision integral $C[f]$ appearing in
Eq.~\eqref{eq:boltz_cyl} is computed in the 
Anderson-Witting relaxation time approximation 
(RTA) \cite{Anderson:1974,Anderson:1974b},
\begin{equation}
 C[f] \rightarrow C_{\rm A-W}[f] = 
 -\frac{k \cdot u}{\tau_{\rm R}} 
 [f - f^{\rm (eq)}],\label{eq:AW}
\end{equation}
where $\tau_{\rm R}$ is the relaxation time and 
$f^{\rm (eq)}$ is the local equilibrium distribution
function. In this paper, we consider that the
equilibrium statistics are described by the 
Maxwell-J\"uttner model for massless particles,
\begin{align}
 f^{\rm (eq)} \rightarrow f^{\rm (eq)}_{\rm M-J} =&
 \frac{g}{(2\pi)^3} 
 \exp\left(\frac{\mu - k \cdot u}{T}\right) \nonumber\\
 =& \frac{n}{8\pi T^3} 
 \exp\left(-\frac{k \cdot u}{T}\right),
\end{align}
where $g$ is a degeneracy factor ($g = 16$ for the 
gluonic degrees of freedom), while $\mu$ and $T$ are 
the local chemical potential and temperature,
respectively. The macroscopic velocity 
$u = u^\halpha e_\halpha$ is obtained via the 
Landau matching condition,
\begin{equation}
 T^\halpha{}_\hsigma u^\hsigma = e u^\halpha,
 \label{eq:Landau_def}
\end{equation}
where the energy density $e = 3p$ represents the
positive eigenvalue of the stress-energy tensor,
$T^\halpha{}_\hsigma$. The temperature $T = p / n$ 
is determined using the particle number density $n$,
which is computed from the particle four-flow and is
related to the chemical potential $\mu$ via
\begin{equation}
 n = N^\halpha u_\halpha = 
 \frac{g T^3}{\pi^2} e^{\mu / T}.
\end{equation}
Both the stress-energy tensor and the particle four-flow
are computed using Eq.~\eqref{eq:macro_vielb} from the
distribution function $f$. 

In the case of the $0+1$D Bjorken flow, there is no
dependence on the spatial coordinates $r$, $\theta$ and
$\eta_s$, while the macroscopic velocity is given by
$u^\halpha = (1,0,0,0)^T$ at all times. Since we
consider no dependence on the azimuthal coordinate 
$\varphi$ of the momentum space in the initial
Romatschke-Strickland distribution given in
Eq.~\eqref{eq:RS}, it is clear that 
$\partial_\varphi f = 0$ at all times and
Eq.~\eqref{eq:boltz_gen} reduces after setting 
$m = 0$ and $k^\htau = k$ to 
\begin{multline}
 \frac{1}{\tau} \frac{\partial (f\tau)}{\partial \tau} 
 - \frac{\xi^2}{\tau k^2} 
 \frac{\partial(f k^3)}{\partial k} - \frac{1}{\tau} 
 \frac{\partial[\xi(1 - \xi^2) f]}{\partial \xi} \\
 = -\frac{1}{\tau_R}[f - f^{\rm (eq)}].
 \label{eq:boltz_0p1D}
\end{multline}
In the free-streaming limit, $\tau_R \rightarrow \infty$
and the solution of Eq.~\eqref{eq:boltz_0p1D} is given
at time $\tau > \tau_0$ precisely by
\begin{equation}
 f_{\rm FS}(\tau; k,\xi) = 
 f_{\rm FS}(\tau_0; \widetilde{k}, \widetilde{\xi}),
 \label{eq:1D_FS}
\end{equation}
where \cite{Kurkela:2018qeb}
\begin{equation}
 \widetilde{k} = k \sqrt{1 - 
 \left(\frac{\tau^2}{\tau_0^2} - 1\right) \xi^2}, \qquad 
 \widetilde{\xi} = \frac{k \xi}{\widetilde{k}}
 \frac{\tau}{\tau_0}.
\end{equation}
Assuming that the distribution at initial time 
$f_{\rm FS}(\tau_0; k, \xi)$ is given by the 
Romatschke-Strickland $f_{\rm RS}(\tau_0; k, \xi)$ distribution in Eq.~\eqref{eq:RS}, the 
free-streaming solution \eqref{eq:1D_FS} reduces 
to Eq.~\eqref{eq:FS_bjork} in the main text.

In the case with transverse expansion, the longitudinal 
boost invariance and the invariance under azimuthal 
plane rotations imply that 
$\partial_\theta f = \partial_{\eta_s} f = 0$.
Restricting the discussion to massless particles, 
when $k^\htau = k$, Eq.~\eqref{eq:boltz} reduces to
\begin{multline}
 \frac{1}{\tau} \frac{\partial (f\tau)}{\partial \tau} + 
 \frac{k^\hatr}{r k} \frac{\partial (f r)}{\partial r} 
 - \frac{\xi^2}{\tau k^2} 
 \frac{\partial(f k^3)}{\partial k} \\
 - \frac{1}{\tau} 
 \frac{\partial[\xi(1 - \xi^2) f]}{\partial \xi} - 
 \frac{\sqrt{1 - \xi^2}}{r} 
 \frac{\partial(f \sin\varphi)}{\partial \varphi} \\ 
 = -\frac{k \cdot u}{k \tau_R} [f - f^{\rm (eq)}].
 \label{eq:boltz_cyl}
\end{multline}
The Landau frame velocity 
$u^\halpha = (u^\htau, u^\hatr, 0,0)$ and the 
energy density $e$ are given by the solution of 
the eigenvalue equation 
\eqref{eq:Landau_def} \cite{Ambrus:2018kug},
\begin{align}
 e =& \frac{1}{2}\left[T^{\htau\htau} - T^{\hatr\hatr} + 
 \sqrt{(T^{\htau\htau}+T^{\hatr\hatr})^2 -
 4(T^{\htau\hatr})^2}\right],\nonumber\\
 \frac{u^\hatr}{u^\htau} =& 
 \frac{T^{\htau\hatr}}{e + T^{\hatr\hatr}}.
\end{align}

\subsection{Momentum space discretization}\label{app:RTA:discretization}

In this paper, we employ the discretization of the
momentum space discussed in Ref.~\cite{Ambrus:2018kug}.
In this scheme, we employ 
$Q_L \times Q_\xi \times Q_\varphi$ discrete values 
for $k$, $\xi$ and $\varphi$, such that 
$k^\halpha \rightarrow k_{l j i}^\halpha = 
k_l (1, \sqrt{1 - \xi_j^2} \cos\varphi_i, 
\sqrt{1 - \xi_j^2} \sin\varphi_i, \xi_j)$. 
The discrete set of distributions $f_{l j i}$ are
related to the original distribution function 
$f(k,\xi,\varphi)$ via \cite{Ambrus:2018kug}
\begin{equation}
 f_{l j i} = \frac{2\pi}{Q_\varphi} T_0^3 e^{\wk_l}
 w_l^L w_j^\xi f(k_l, \xi_j, \varphi_i).
\end{equation}
The weights $w_l^L$, $w_j^\xi$ and $2\pi / Q_\varphi$
are computed following the prescription of the 
Gauss-Laguerre, Gauss-Legendre and Mysovskikh
(trigonometric) quadratures, respectively \cite{Mysovskikh:1987}.

The values for $k$ are chosen as the roots of the
generalized Laguerre polynomials $L_{Q_L}^{(2)}(\wk)$ 
of order $Q_L$, where $\wk = k / k_{\rm ref}$ and
$k_{\rm ref}$ is an arbitrary scale which we set equal
to the initial temperature, $k_{\rm ref} = T_0$. Two
values are chosen ($Q_L = 2$), $\wk_1 = 2$ and 
$\wk_2 = 6$, thus ensuring the exact recovery of 
the evolution of $N^\halpha$ and $T^{\halpha\hsigma}$
(for details, see Ref.~\cite{Ambrus:2018kug}). 
The corresponding weights are $w_1^L = 3/2$ and 
$w_2^L = 1/2$. The derivative term 
$k^{-2} \partial (fk^3) /\partial k$ is projected onto
the space of generalized Laguerre polynomials and is
truncated at order $Q_L$, giving
\begin{equation}
 \left[\frac{1}{k^2} \frac{\partial (f k^3)}{\partial k}
 \right]_{lji} = \sum_{\ell' = 1}^{Q_L} 
 \mathcal{K}^L_{l,l'} f_{l' ji},
\end{equation}
where the elements of the $Q_L \times Q_L$ matrix 
$\mathcal{K}^L_{l,l'}$ are given in Eq.~(3.51) 
of Ref.~\cite{Ambrus:2018kug}. For $L= 2$, 
$\mathcal{K}^L_{l,l'} = \frac{1}{6} 
w_l^L (3 - \wk_l) \wk_{l'}$, such that 
$\mathcal{K}^L_{1,1'} = -\mathcal{K}^L_{2,1'} = 
\frac{1}{2}$ and 
$\mathcal{K}^L_{1,2'} = -\mathcal{K}^L_{2,2'} = 
\frac{3}{2}$, leading to
\begin{align}
 \left[\frac{1}{k^2} \frac{\partial (f k^3)}{\partial k}
 \right]_{1ji} =& \frac{1}{2} f_{1ji} + 
 \frac{3}{2} f_{2ji}, \nonumber\\
 \left[\frac{1}{k^2} \frac{\partial (f k^3)}{\partial k}
 \right]_{2ji} =& -\frac{1}{2} f_{1ji} - 
 \frac{3}{2} f_{2ji}.
\end{align}

In the case of $\xi$, we employ the Gauss-Legendre
quadrature of order $Q_\xi$, meaning that the values 
$\xi_j$ ($1 \le j \le Q_\xi$) are the roots of the
Legendre polynomial of order $Q_\xi$, 
$P_{Q_\xi}(\xi_j) = 0$. Both the roots and the weights
$w_j^\xi$ up to order $Q_\xi = 1000$ are available 
as data files in the supplementary material of
Ref.~\cite{Ambrus:2018kug}. The term 
$\partial [\xi(1 - \xi^2) f] / \partial \xi$ is 
computed by projection onto the space of Legendre
polynomials,
\begin{equation}
 \left[\frac{\partial[\xi(1 - \xi^2) f]}{\partial \xi
 }\right]_{lji} = \sum_{j' = 1}^{Q_\xi}
 \mathcal{K}^P_{j,j'} f_{lj'i},
\end{equation}
where the $Q_\xi \times Q_\xi$ elements of the matrix 
$\mathcal{K}^P_{j,j'}$ are computed from Eq.~(3.54) 
of Ref.~\cite{Ambrus:2018kug}, reproduced below for
convenience:
\begin{multline}
 \mathcal{K}^P_{j,j'} = w_j \sum_{m = 1}^{Q_\xi - 3} 
 \frac{m(m+1)(m+2)}{2(2m+3)} P_m(\xi_j) 
 P_{m+2}(\xi_{j'}) \\
 - w_j \sum_{m = 1}^{Q_\xi - 1} \frac{m(m+1)}{2}
 P_m(\xi_j) \Bigg[\frac{(2m + 1) 
 P_m(\xi_{j'})}{(2m-1)(2m+3)}\\
 + \frac{m-1}{2m-1} P_{m-2}(\xi_{j'})\Bigg].
\end{multline}

Finally, the trigonometric angle $\varphi$ is
discretized using $Q_\varphi$ values, 
$\varphi_i = \varphi_0 + 2\pi (i-1) / Q_\varphi$ 
($1 \le i \le Q_\varphi$), where the arbitrary offset 
$\varphi_0$ is set to $\varphi_0 = 0$ for definiteness.
The derivative term 
$\partial (f \sin\varphi) / \partial \varphi$ can 
be computed via
\begin{equation}
 \left[\frac{\partial(f \sin\varphi)}{\partial \varphi}
 \right]_{lji} = \sum_{i' = 1}^{Q_\varphi} 
 \mathcal{K}^M_{i,i'} f_{l,j,i'},
\end{equation}
where
\begin{multline}
 \mathcal{K}^M_{i,i'} = \frac{1}{Q_\varphi} 
 \sum_{m = 0}^{\lfloor Q_\varphi / 2 \rfloor} \{
 (m+1) \cos[m(\varphi_i - \varphi_{i'}) + \varphi_i] \\
 - (m -1) \cos[m(\varphi_i - \varphi_{i'}) - \varphi_i]
 \},
\end{multline}
where $\lfloor \cdot \rfloor$ is the floor function.

Before ending this subsection, we discuss the strategy
employed for computing the initial conditions for $f$,
as well as the equilibrium distribution $f^{\rm (eq)}$.
Because the momentum magnitude $k$ is discretized using
only two values, the direct evaluation of the
distribution function at these values suffers from
severe accuracy problems when attempting to extract 
the macroscopic quantities $N^\halpha$ and
$T^{\halpha\hbeta}$. Instead, we employ the strategy of
Refs.~\cite{Romatschke:2011hm,Ambrus:2018kug} and
consider the projection of $f$ onto the space of
Laguerre polynomials:
\begin{align}
 f =& \frac{e^{-k / T_0}}{T_0^3} 
 \sum_{\ell = 0}^{Q_L - 1}
 \frac{\mathcal{F}_\ell L_\ell^{(2)}(k/T_0)}
 {(\ell+1)(\ell+2)}, \nonumber\\
 \mathcal{F}_\ell =& \int_0^\infty \d p\,p^2\, f\,
 L_\ell^{(2)}(k/T_0).
\end{align}
The sum over $\ell$ is truncated at $Q_L - 1$ in order 
to facilitate the recovery of the integrals of $f$ 
following the Gauss-Laguerre quadrature prescription.
The initialization of $f$ is performed at the level of
the coefficients $\mathcal{F}_\ell$, which for the
Romatschke-Strickland distribution are given by
\cite{Ambrus:2019wfc}:
\begin{align}
 \mathcal{F}^{\rm RS}_0 =& 
 \frac{g e^{\alpha_0}}{(2\pi)^3} 
 \frac{\Lambda_0^3}{(1 + \xi_0 \xi^2)^{3/2}},
 \nonumber\\
 3\mathcal{F}^{\rm RS}_0 - \mathcal{F}^{\rm RS}_1 =& 
 \frac{3g e^{\alpha_0}}{(2\pi)^3} 
 \frac{\Lambda_0^4 / T_0}{(1 + \xi_0 \xi^2)^2}.
\end{align}
After the discretization of the momentum space, 
the Romatschke-Strickland distribution becomes
\begin{multline}
 f^{\rm RS}_{lji} = 
 \frac{g e^{\alpha_0} \Lambda_0^3}
 {(1 + \xi_0 \xi_j^2)^{3/2}} 
 \frac{w_l^L w_j^\xi}{4\pi^2 Q_\varphi} \\\times
 \left[4 - {\bar k}_l + 
 \frac{\Lambda_0 / T_0}{\sqrt{1 + \xi_0 \xi_j^2}} 
 (3 - {\bar k}_l)\right].
\end{multline}
The Maxwell-J\"uttner distribution necessary for 
the computation of the collision term is obtained by
replacing $e^{\alpha_0} \rightarrow \pi^2 n / g T^3$, 
$\Lambda_0 \rightarrow T$ and 
$\sqrt{1 + \xi_0 \xi^2} \rightarrow 
u_\halpha k^\halpha / k = 
u^\htau - k^\hatr u^\hatr / k$:
\begin{multline}
 f^{\rm {(eq)}}_{lji} = 
 \frac{n w_l^L w_j^\xi}{4 Q_\varphi (u^\htau - 
 u^\hatr \sqrt{1 - \xi_j^2} \cos\varphi_i)^3} \\\times
 \left[4 - {\bar k}_l 
 + \frac{(3 - {\bar k}_l) T / T_0}
 {u^\htau - u^\hatr \sqrt{1 - \xi_j^2} \cos\varphi_i}
 \right].
\end{multline}

\subsection{Finite difference methods}
\label{app:RTA:numsch}

The time stepping is performed using the third-order
Runge-Kutta scheme \cite{Shu:1988,Gottlieb:1998}. 
Writing Eq.~\eqref{eq:boltz_cyl} as 
$\partial f / \partial \tau = L[f]$ and considering 
an equal time step discretization of the time
coordinate, $\tau_n = \tau_0 + n \delta \tau$, the value
$f_{n + 1}$ of the distribution function at time step 
$n + 1$ can be obtained from that at time step $n$ via
two intermediate stages:
\begin{align}
 f^{(1)}_n =& f_n + \delta t L[f_n], \nonumber\\
 f^{(2)}_n =& \frac{3}{4} f_n + \frac{1}{4} f^{(1)}_n 
 + \frac{1}{4} \delta t L[f^{(1)}_n], \nonumber\\
 f_{n+1} =& \frac{1}{3} f_n + \frac{2}{3} f_n^{(2)} 
 + \frac{2}{3} \delta t L[f^{(2)}_n].
\end{align}

For the advection along $r$, care must be taken 
because of the $r^{-1}$ factor appearing in
Eq.~\eqref{eq:boltz_cyl}. Following 
Refs.~\cite{Falle:1996,Downes:2002,Busuioc:2019}, 
this factor is absorbed in the derivative, i.e. 
$r^{-1} \partial(f r) / \partial r = 
2 \partial(f r) / \partial r^2$. The discretization 
of the radial coordinate is performed using $S$ equal
intervals of width $\delta r = L / S$ (where $L = 6w$
and $w$ is the width of the Gaussian, as discussed in
Sec.~\ref{sec:transv}), centered on coordinates 
$r_s = (s - \frac{1}{2}) \delta r$, with 
$1 \le s \le S$. The derivative term is then computed
using a flux-based finite-difference scheme,
\begin{equation}
 \left[\frac{1}{r} \frac{\partial(f v r)}{\partial r}
 \right]_s = 
 2 \frac{r_{s + \frac{1}{2}} \mathcal{F}_{s+\frac{1}{2}}
 - r_{s - \frac{1}{2}} \mathcal{F}_{s - \frac{1}{2}}}
 {r_{s +\frac{1}{2}}^2 - r_{s-\frac{1}{2}}^2},
 \label{eq:advection}
\end{equation}
where $r_{s \pm \frac{1}{2}} = 
r_s \pm \frac{1}{2} \delta r$ and 
$v = k^\hatr / k^\htau \rightarrow 
\cos\varphi_i \sqrt{1 - \xi_j^2}$ is the advection 
velocity. The fluxes $\mathcal{F}_{s \pm \frac{1}{2}}$
are computed using an upwind-biased approach. For
increased stability, we employ the fifth-order weighted
essentially nonoscillatory (WENO-5) scheme 
\cite{Jiang:1996,Rezzolla:2013}, which is summarized
also in Refs.~\cite{Ambrus:2018kug,Busuioc:2019}. 
For brevity, the algorithm is not repeated here. 
We note that, while the WENO-5 method is of fifth order,
the formulation in Eq.~\eqref{eq:advection} gives rise
to a second-order algorithm due to the 
$r_{s \pm \frac{1}{2}}$ factors appearing in 
the numerator \cite{Busuioc:2019}. 

\bibliography{etas}

\end{document}